\documentclass[aip, cha, reprint]{revtex4-1}

\usepackage{graphicx}
\usepackage{amsmath,amsfonts,amsthm,amssymb}

\usepackage{tikz}

\usepackage{color}
\usepackage{epstopdf}

\usepackage{hyperref}
\hypersetup{
    colorlinks,
    citecolor=blue,
    filecolor=blue,
    linkcolor=blue,
    urlcolor=blue
}

\newcommand{\refcitenosup}[1]{\setcitestyle{numbers}{Ref.~\cite{#1}}\setcitestyle{super}}

\newcommand{\Mpp}{\ensuremath M_2}
\newcommand{\Mppinv}{\ensuremath M^{-1}_2}

\newcommand{\Mps}{\ensuremath M_2}

\newcommand{\Mapiii}{\ensuremath M_3}
\newcommand{\Mapiiiinv}{\ensuremath M^{-1}_3}
\newcommand{\Maciii}{\ensuremath M_3}
\newcommand{\Maciiiinv}{\ensuremath M^{-1}_3}

\newcommand{\Maciiirevburn}{\ensuremath\mathcal{R}}

\newcommand{\Mapii}{\ensuremath\mathcal{M}}

\newcommand{\Masii}{\ensuremath\mathcal{M}}

\newcommand{\Masiiloose}{\ensuremath\overline{\mathcal{M}}^{-1}}
\newcommand{\Masiitight}{\ensuremath\underline{\mathcal{M}}^{-1}}

\definecolor{johnred}{rgb}{0.8, 0.2, 0.2}
\definecolor{kevingreen}{rgb}{0.2, 0.8, 0.2}

\newcommand{\JM}[1]{\textbf{\textcolor{johnred}{#1}}}

\begin{document}

\title{A turnstile mechanism for fronts propagating in fluid flows}
\author{John~R.~Mahoney}
\affiliation{School of Natural Sciences, University of California, Merced, CA 95344, USA}
\author{Kevin~A.~Mitchell}
\affiliation{School of Natural Sciences, University of California, Merced, CA 95344, USA}
\date{\today}

\begin{abstract}
We consider the propagation of fronts in a periodically driven flowing medium. It is shown that the progress of fronts in these systems may be mediated by a turnstile mechanism akin to that found in chaotic advection.
We first define the modified (``active'') turnstile lobes according to the evolution of point sources across a transport boundary.
We then show that the lobe boundaries may be constructed from stable and unstable \emph{burning invariant manifolds}---one-way barriers to front propagation analogous to traditional invariant manifolds for passive advection.
Because the burning invariant manifolds (BIMs) are one-dimensional curves in a three-dimensional ($xy\theta$) phase space, their projection into $xy$-space exhibits several key differences from their advective counterparts:
(lobe) areas are not preserved, BIMs may self-intersect, and an intersection between stable and unstable BIMs does not map to another such intersection.
These differences must be accommodated in the correct construction of the new turnstile. 
As an application, we consider a lobe-based treatment protocol for protecting an ocean bay from an invading algae bloom.
\end{abstract}


\maketitle
\label{sec:lead_paragraph}
\begin{quotation}
Fronts in flowing media are the combination of two complementary physical phenomena. 
Front propagation is governed by geometric evolution, built around the structure of rays and perpendicular fronts.
Fluid flows exhibit structures that are sinewy, tenuous, and generically chaotic even for simple flows.
Here we study the interplay of these two dynamics.

A central concept in both front propagation and fluid flows is that of \emph{transport}---moving something from here to there. A laser pulse traveling through fiber optic cable transports information, while an ocean wave transports energy and surfers. 
While some instances of fluid transport have a similar linear character, e.g., the oceanic conveyor belt that moves heat around the globe, we will be more interested in fluid transport as a means to mixing, a process that depends on the chaotic nature of the flow.

This study examines how this chaotic transport in fluids is augmented by propagation.
Our approach draws from the existing geometric theory of transport in fluid flows known as the \emph{turnstile mechanism}. 
This mechanism is a first-order description of chaotic mixing which describes the exchange of fluid packets between two adjacent regions.
We develop a reformulation of this theory that naturally encompasses the addition of front propagation.
\end{quotation}

\section{Introduction}
\label{sec:introduction}
We build upon the foundation subjects of fluid flow and front propagation, both central dynamical mechanisms in Nature.
Exemplars of fluid flow include oceanic and atmospheric flow, industrial mixing, and microfluidics.
We might also include such systems as phase space flow and swarms of moving agents.
Well-studied cases of front propagation include optical and acoustic wavefronts, phase transition fronts, and chemical reaction fronts.
Again we can broaden our view to include growth of algae blooms, spread of infectious disease, and dissemination of information and beliefs.

It is often the case that the substrate supporting the propagation of a front is not static, and certainly may be some form of flowing fluid.
The combination of these two elemental dynamics yields a variety of new and interesting phenomena while maintaining features of each component.

Recent experiments have combined chemical reaction front propagation with laboratory scale flows \cite{Paoletti05, Schwartz08, Bargteil12, Pocheau06, Pocheau08}.
The flows were generated using magneto-hydrodynamic forcing in a quasi-two-dimensional fluid.
The reaction fronts were generated through Belousov-Zhabotinsky chemistry.
A variety of phenomena were investigated including mode-locking and front-pinning. Direct numerical simulations \cite{Abel01, Cencini03} have supported these experiments.

Here we take a different perspective.
Invariant manifolds are key to understanding passive advection in time-independent and time-periodic fluid flows \cite{Wiggins92}.
Recent experiment and theory \cite{Bargteil12, Mahoney12, Mitchell12b} has shown that these invariant manifolds survive, but are modified by, the addition of front propagation to the dynamics. The original (advective) invariant manifolds undergo a bifurcation whereby they split into two nearby \emph{oriented} invariant manifolds, called \emph{burning invariant manifolds} (BIMs). Such manifolds are oriented in that they are \emph{one-sided} barriers to the progress of fronts in fluid flows.

The fundamental geometric construct for understanding chaotic transport in passive flows is the turnstile mechanism \cite{MacKay84}. The turnstile is built from segments of advective stable and unstable invariant manifolds. Since BIMs are relevant for front propagation, this suggests that a turnstile-type mechanism may also be at work in active fluid flows. 

We begin in Section~\ref{sec:turnstile_review} with a review of the turnstile mechanism in passive advection.
Then Section~\ref{sec:dynamical_system} presents the dynamical system for fronts in flows.
Section~\ref{sec:burnstile} defines the new burning lobes in terms of point stimulations.
We then recall the necessary features of BIMs from previous works (Section~\ref{sec:BIM_basics}).
Sections~\ref{sec:small_v_0} and \ref{sec:large_v_0} examine the modifications necessary for our new turnstile given small and large front propagation speeds, respectively.
We end by using this burning turnstile structure to prescribe an efficient treatment of an oceanic algae bloom (Section~\ref{sec:reaction_suppression}).

\section{advective turnstile review}
\label{sec:turnstile_review}

\begin{figure}
\includegraphics[width=\linewidth]{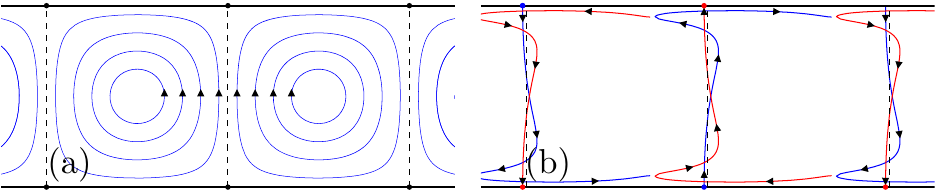}
\caption{Structure in alternating-vortex-chain fluid flow. (a) Time-independent flow forms vortex cells bounded by separatrices (dashed lines). $(b=0, \omega=2 \pi, \phi=0)$ (b) Periodic lateral perturbation breaks each separatrix into stable (red) and unstable (blue) manifolds. $(b=0.1, \omega=2 \pi, \phi=0)$}
\label{fig:passive_broken_separatrix}
\end{figure}

Time-independent, two-dimensional, incompressible flows are well characterized by the structure of their fixed points and separatrices. The separatrices are invariant curves under the flow. They divide the fluid trajectories into classes with qualitatively distinct behaviors (Fig.~\ref{fig:passive_broken_separatrix}(a)). Unlike the invariant tori that are also present in these flows, separatrices divide trajectories that exponentially diverge or converge.

The introduction of a time-periodic perturbation leads to the splitting of each separatrix into distinct stable and unstable invariant manifolds (Fig.~\ref{fig:passive_broken_separatrix}(b)). These manifolds are not invariant with respect to the time-varying flow, but rather the map $\Mpp: \mathbb{R}^2 \to \mathbb{R}^2$ obtained by integrating the flow over one forcing cycle.

As a model system for illustrating these structures, we use the alternating vortex chain (AVC) in a channel geometry \cite{Solomon88},
\begin{equation}
\begin{aligned}
\label{eq:velfield}
u_x(x, y, t) &= + \sin(\pi[x + b \sin (\omega t - \phi)]) \cos(\pi y ),\\
u_y(x, y, t) &= - \cos(\pi[x + b \sin (\omega t - \phi)]) \sin(\pi y ),
\end{aligned}
\end{equation}
where $0 \le y \le 1$, and $b$, $\omega$, and $\phi$ are dimensionless parameters that describe the periodic forcing amplitude, frequency, and phase, respectively. (We use the phase $\phi$ to choose a particular Poincar\'e section.) This model has free-slip boundary conditions (BCs).
We will use this same model in following sections when discussing active fluids.

For small perturbations and short times, the unforced separatrices are still fairly good classifiers of particle trajectories. However, these separatrices, are traversed by certain trajectories. This is sometimes referred to as a ``leaky separatrix''. 
Importantly, this is not ``leakiness'' in the sense of diffusion. Rather, there is structure to the leakiness. One iteration of the map causes a localized packet of fluid to be transported across from left to right, and a corresponding packet from right to left. 

\subsection{Set definition of lobes}
To formalize these ideas, we first choose a transport boundary that divides the fluid in two allowing us to define a flux. In this paper, we focus on flow in a channel geometry and choose a transport boundary that defines a left- and right-hand side ($LHS$, $RHS$).(Starting in Sec.~\ref{sec:burnstile}, we specialize to transport from left to right). Given this transport boundary, we define lobes as sets of points that cross this boundary from left to right, or right to left, upon one iteration of the map $\Mpp$.
The two sets, before and after the map, are referred to as lobes. The escape lobes $E_i$ and capture lobes $C_i$ are defined as
%
\begin{equation}
\begin{aligned}
\label{eqn:passive_lobe_defs_pts}
E_{0} &= \{p \in RHS : \exists q \in LHS, \Mpp(q) = p\},\\
E_{-1} &= \{p \in LHS : \Mpp(p) \in RHS\},\\
C_{0} &= \{p \in LHS : \exists q \in RHS, \Mpp(q) = p\},\\
C_{-1} &= \{p \in RHS : \Mpp(p) \in LHS\}.
\end{aligned}
\end{equation}
We use the same notation to denote the map applied to a set of points  $\Mps(A) = \{\Mpp(p): p \in A\}$. 
Using this notation and the invertibility of $\Mpp$, it is somewhat more intuitive to write,
\begin{equation}
\begin{aligned}
\label{eqn:passive_lobe_defs_sets}
E_{0} &= RHS \cap \Mps(LHS),\\
E_{-1} &= \Mps^{-1}(RHS) \cap LHS,\\
C_{0} &= LHS \cap \Mps(RHS),\\
C_{-1} &= \Mps^{-1}(LHS) \cap RHS,
\end{aligned}
\end{equation}
%
noting that $E_{0} = \Mps(E_{-1})$ and $C_{0} = \Mps(C_{-1})$.

\begin{figure}
\includegraphics[width=\linewidth]{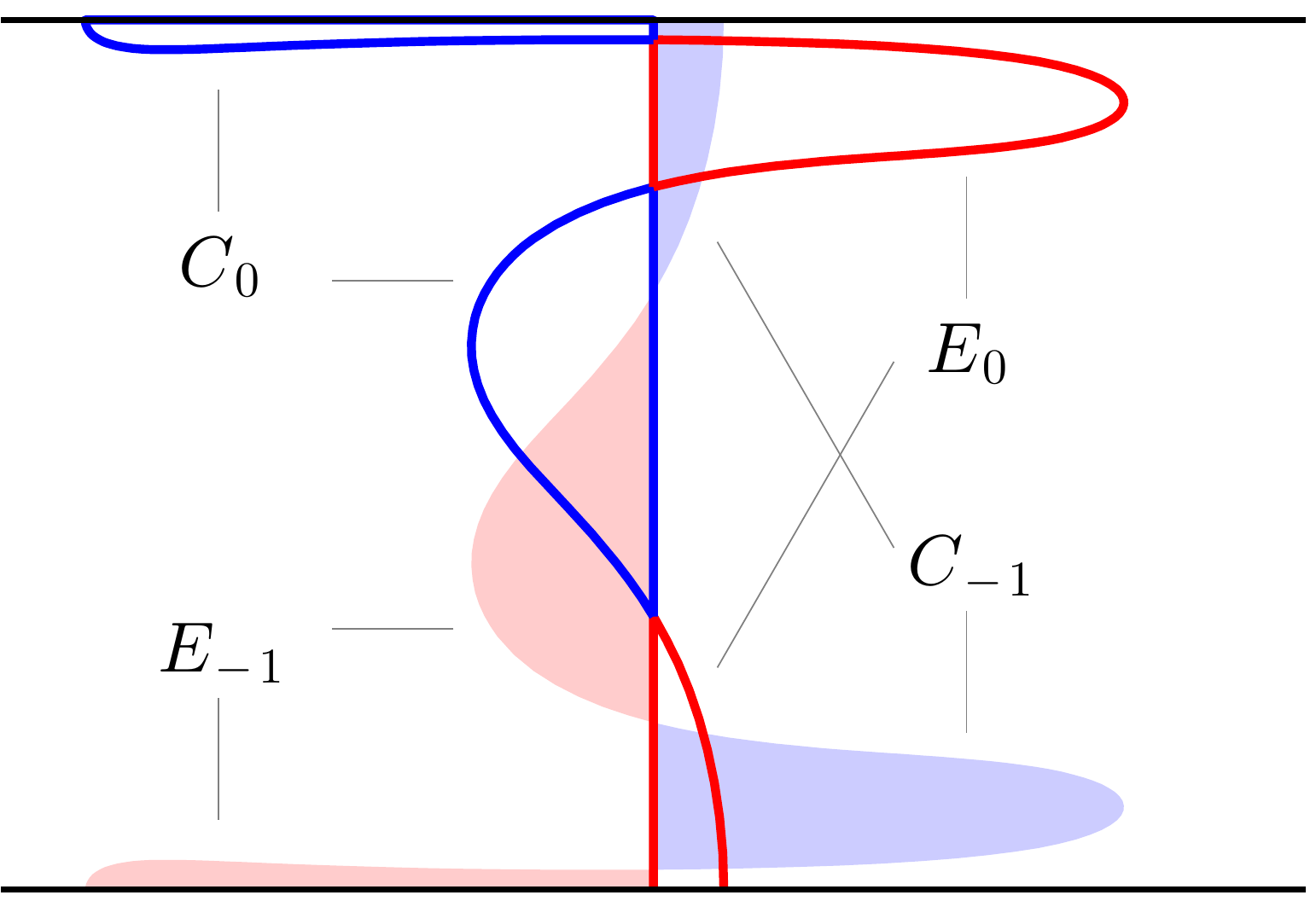}
\caption{Separatrix (central vertical line) is chosen as transport boundary for periodically-forced AVC. 
Left to right transport: red filled regions map to red outlined regions ($E_{-1} \to E_0$). Right to left transport: blue filled regions map to blue outlined regions ($C_{-1} \to C_0$). $(b=0.4, \omega=5, \phi=\pi)$}
\label{fig:arbitrary_transport_boundary}
\end{figure}

In Figure~\ref{fig:arbitrary_transport_boundary}, we choose the advective separatrix as our transport boundary (vertical line in the middle). Examining the action of one map iteration, we find eight regions of interest. Two disjoint regions map to two other disjoint regions going left to right ($E_{-1} \to E_0$), and similarly going right to left ($C_{-1} \to C_0$). Notice that these regions are bounded both by segments of the separatrix and also by segments of its forward and reverse iterates, which do not coincide with the separatrix. Also, note the overlapping of regions and the amount of total area involved. While this analysis of single-step transport is correct, it can be simplified by a different choice of transport boundary.

\subsection{Transport boundary - invariant manifolds}
While the above definition yields well-defined lobes given \emph{any} transport boundary, it is standard practice to construct this boundary from segments of stable and unstable invariant manifolds \cite{Wiggins92}. In doing so, we describe the transport process using ``native'' structures. Two hyperbolic fixed points (on opposite sides of the channel) are denoted $z^A$ and $z^B$. The stable and unstable manifolds attached to these points are denoted $W^S$ and $W^U$, respectively. The finite segments connecting these fixed points to the intersection point $p_0$ are denoted $W^S[z^A, p_0]$ and $W^U[z^B, p_0]$. The intersection point is required to be a \emph{primary intersection point} \cite{Rom-Kedar90b}; that is, $W^S[z^A, p_0] \cap W^U[z^B, p_0] \ \backslash \ \{z^A, z^B, p_0\}=\emptyset$. The transport boundary is then defined as $W^S[z^A, p_0] \cup W^U[z^B, p_0]$. 

\begin{figure}
\includegraphics[width=\linewidth]{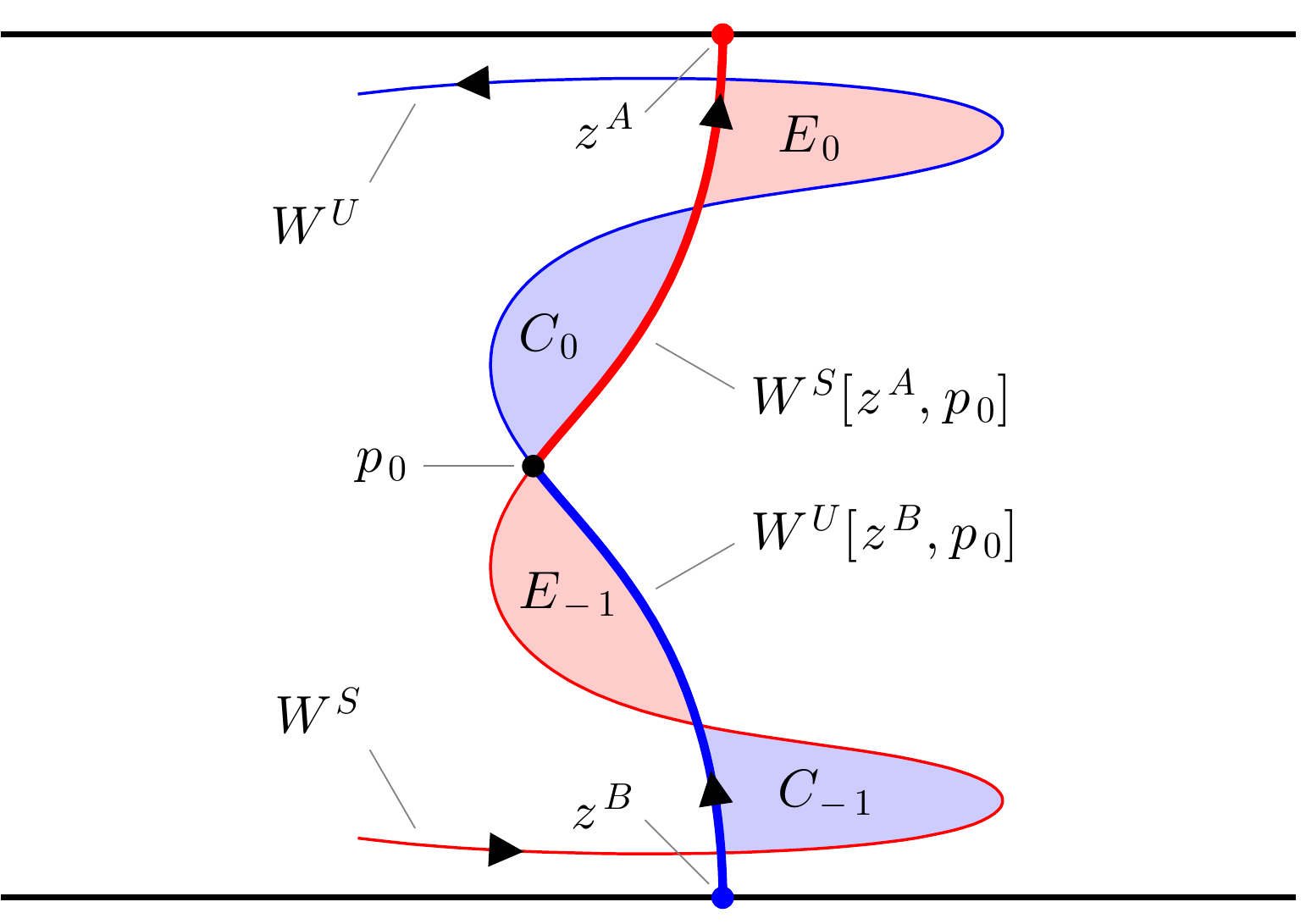}
\caption{Turnstile mechanism in periodically forced alternating vortex chain. Segments of stable (red) and unstable (blue) manifold are combined to form transport boundary (bold). Shaded areas indicate the two pathways for crossing this transport boundary. Use of dynamically generated transport boundary yields simpler lobe structure. $(b=0.4, \omega=5, \phi=\pi)$}
\label{fig:advective_turnstile}
\end{figure}

In Fig.~\ref{fig:advective_turnstile}, we illustrate the lobes derived from this new transport boundary. Importantly, these lobe boundaries are comprised solely of segments of these same invariant manifolds.
Each lobe is also composed of a smaller number (here just one) of disjoint regions having smaller total area.

The $C_{-1}$ and $E_{-1}$ lobes share a point in their boundary and, loosely speaking, the complementary transport of these packets corresponds to a rotation (and shift) about this point. Consequently, this exchange is referred to as a \emph{turnstile mechanism} \cite{MacKay84} \footnote{It should be pointed out that the turnstile mechanism is a piece of the larger theory of lobe dynamics. 
The goal of this paper is to firmly establish the turnstile mechanism for propagating fronts. A treatment of the multi-time-step dynamics is left to a future paper.}. 
Generalizing the connection between the evolution of sets and bounding invariant manifolds to the case of front propagation is one of the main results of this paper.
%

\section{A dynamical system for fronts in flows}
\label{sec:dynamical_system}
\subsection{Assumptions}
\label{sec:assumptions}

We assume that the two-dimensional fluid is partitioned into regions that can be identified as either ``burned'' or ``unburned''---we use ``burned'' generically to refer to the region bounded by an outward propagating front. That is, these regions are separated by well-defined one-dimensional (codimension one) curves or \emph{fronts}. This is known as the ``thin (sharp) front'', or ``geometric optics'' regime \cite{Ronney94}. For instance, if our front is an autocatalytic chemical reaction front, this limit requires that the diffusion rate is small compared with the reaction rate.

We assume that each front element (infinitessimal piece of the one-dimensional front) propagates perpendicular to the front at a speed $v_0$ measured in the comoving fluid frame, thereby expanding the burned region. To be clear, we require that this speed $v_0$ be isotropic, homogeneous, and time-independent.
Note that we have implicitly assumed that the front elements are non-interacting (fronts do not change their behavior when faced with a collision), and their motion does not depend on the front curvature.

Finally, we assume that there is no effect of the front propagation on the underlying fluid flow. In the case of a chemical reaction front, this means that the reaction is not sufficient to change the fluid density, viscosity, etc. This is a good approximation in such experiments as \refcitenosup{Paoletti05, Schwartz08, Bargteil12}.

\subsection{Reacted region PDE}

The evolution of a reacted region can be modeled using the modified FKPP equation \cite{Fisher37, Kolmogorov37},
\begin{align}
\frac{\partial \Theta}{\partial t} = -\mathbf{u} \cdot \nabla \Theta + \frac{1}{\tau} f(\Theta) + D_0 \Delta \Theta,
\label{eqn:FKPP}
\end{align}
where $\phi$ is the local fraction of reaction products, $\tau$ is the reaction time-scale, $f$ describes the reaction kinetics, and $D_0$ is the diffusion constant.
We require the sharp front limit of this equation so that the fronts dividing burned from unburned regions are well defined.

We define the map $\Masii: \mathbb{R}^2 \to 2^{\mathbb{R}^2}$, which takes a point stimulation in $\mathbb{R}^2$ and returns the reacted domain obtained by integrating Eq.~(\ref{eqn:FKPP}) (in the sharp front limit) for one forcing period. Here $2^{\mathbb{R}^2}$ is the collection of subsets of $\mathbb{R}^2$.
The reaction dynamics propagates this point stimulation into some finite region that is all the while stretched and folded by the flow dynamics.


We extend the map $\Masii$ to act on regions by defining $\Masii: 2^{\mathbb{R}^2} \to 2^{\mathbb{R}^2}$ such that $\Masii(A) = \cup\{\Mapii(p): p \in A\}$. The intuition is simply that if stimulation $p_1$ evolves to region $R_1$ and $p_2$ to $R_2$, then stimulating both $p_1$ and $p_2$ will result in the reacted domain $R_1 \cup R_2$. 
Note that $\Masii$ acting on sets is not one-to-one, i.e., two distinct burned regions can evolve forward to the same burned region. This differs from the passive case in which $\Mpp$ acting on sets was invertible (Sec.~\ref{sec:turnstile_review}).

As shown in \refcitenosup{Williams85}, the sharp-front FKPP equation is equivalent to the G-equation,
\begin{align}
\frac{\partial G}{\partial t} = - \mathbf{u} \cdot \nabla G + v_0 |\nabla G|,
\label{eqn:Geqn}
\end{align}
where only the level set $G=0$, and not the value of this field, has any physical meaning. This level set represents the boundary between reacted and unreacted regions. While one can solve this PDE effectively by focusing on the vicinity of the level set, we wish to make an explicit dimension reduction.

\subsection{Front element ODE}

Instead of integrating Eq.~(\ref{eqn:Geqn}), we shall use an equivalent formulation \cite{Oberlack10} in which each front element evolves independently under the following three-dimensional ODE,
\begin{subequations}
\begin{align}
\dot{\mathbf{r}} &= \mathbf{u} + v_0 \hat{\mathbf{n}}, \label{eqn:3DODEa} \\ 
\dot{\theta} &= - \hat{\mathbf{n}}_i \mathbf{u}_{i,j} \hat{\mathbf{g}}_j. \label{eqn:3DODEb}
\end{align}
\label{eqn:3DODE}
\end{subequations}
The variables $\mathbf{r}$ and $\theta$ denote the position and orientation of each front element. The prescribed fluid velocity is given by $\mathbf{u}$. Again, $v_0$ is the front propagation speed in the comoving fluid frame. For compactness, we use the variables $\hat{\mathbf{g}} = (\cos \theta, \sin \theta)$ and $\hat{\mathbf{n}} = (\sin \theta, -\cos \theta)$ to indicate the tangent to the front element and the normal direction (propagation direction), respectively. Finally, $\mathbf{u}_{i,j} = \partial u_i / \partial r_j$ and repeated indices are summed.
This ODE representation allows us to apply the tools of low-dimensional dynamical systems theory.

This ODE can also be derived more intuitively. The total translational motion of a front element is the vector sum fluid velocity plus the front propagation velocity in the fluid frame (Eq.~(\ref{eqn:3DODEa})). The change in orientation is determined entirely geometrically. Equation~(\ref{eqn:3DODEb}) describes the angular velocity of a material line embedded in the fluid.

We define $\Mapiii: \mathbb{R}^2 \times S^1 \to \mathbb{R}^2 \times S^1$, the evolution of front elements, by integrating Eq.~(\ref{eqn:3DODE}) for one forcing period. 
Note that $\Mapiii$ is the active extension of $\Mpp$ (the advection map for a point).
The ``2'' and ``3'' subscripts emphasize that these maps operate on two- and three-dimensional phase spaces.
Finally, we also use $\Maciii(\partial A)$ to denote the iterate of the curve $\partial A$ which is the outward oriented boundary curve (in $xy\theta$-space) of burned region $A$.

\section{Burning turnstile : ``burnstile''}
\label{sec:burnstile}

Here we define the turnstile relevant for active fluids in terms of the behavior of sets, much like was done for the advective turnstile. The essential differences arise due to the everywhere-expanding property of the map $\Masii$.

Because of the asymmetry of the map $\Mapii$ (unburned fluid can become burned but not vice versa) there exists an asymmetry in the lobe definitions depending on whether the burned region is to the left or the right of the transport boundary. For the remainder of this paper, we consider left-to-right propagation. The right-to-left case is defined analogously.

\subsection{Escape}

\begin{figure}
\includegraphics[width = \linewidth]{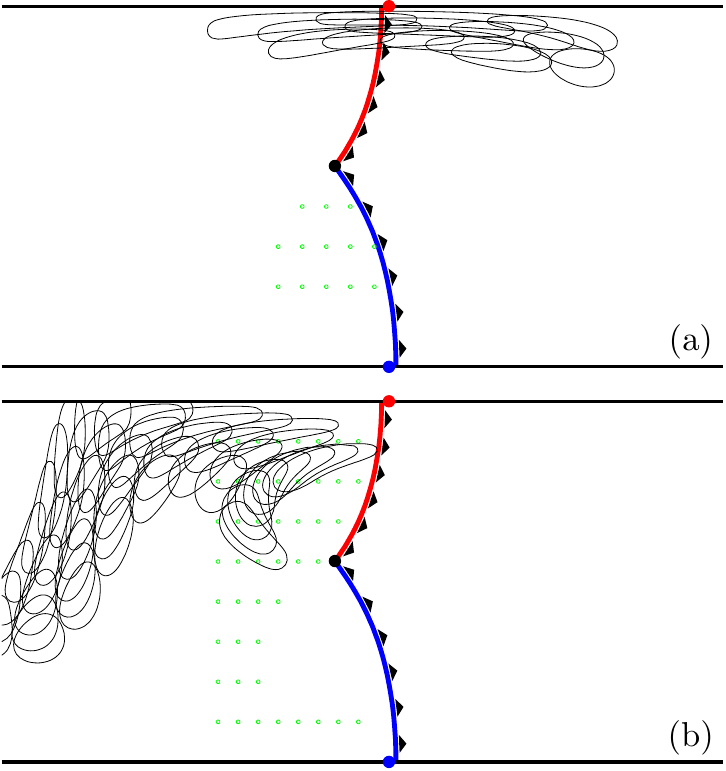}
\caption{Illustration of escape lobe function. Arrows on transport boundary indicate left-to-right choice. (a) Each stimulation evolves such that it intersects the $RHS$ and so must be part of $E_{-1}$. The region covered on $RHS$ is $E_0$. (b) These stimulations remain entirely on the $LHS$ after one iterate and so are not in $E_{-1}$. $(v_0=0.05, b=0.3, \omega=5, \phi=\pi)$}
\label{fig:escape}
\end{figure}

We begin with the escape process. Since advection maps points to points, any given point begins and ends either right or left of the transport boundary, and so crossing (e.g., escaping) is well defined. However with the addition of front propagation, points map to regions under $\Masii$ and so a choice arises; is a point stimulation, which immediately gives rise to a burned region, said to escape when \emph{all} of the region crosses the transport boundary? Or instead when \emph{any} of the region crosses? Looking to physical examples such as autocatalytic chemical reactions, fires, algae blooms, infections, etc., it seems reasonable that the more interesting definition is when \emph{any} crosses; we are generally not concerned whether an \emph{entire} wildfire crosses the firebreak. This motivates the following definitions of escape lobes in the active system (compare Eqs.~(\ref{eqn:passive_lobe_defs_pts})),
\begin{equation}
\begin{aligned}
\label{eqn:active_lobe_defs_ptsE}
E_{-1} &= \{p \in LHS : \Mapii(p) \cap RHS \neq \emptyset\},\\
E_{0} &= \{p \in RHS : \exists p' \in LHS , p \in \Mapii(p')\}.
\end{aligned}
\end{equation}
The escape lobe $E_{-1}$ is the set of all points left of the transport boundary such that the evolution of each point stimulation under $\Masii$ has some intersection with the area right of the transport boundary.
The escape lobe $E_0$ is the set of all points right of the transport boundary that may be reached in one iterate by some point stimulation on the left.
See Fig.~\ref{fig:escape}.
%

\subsection{Capture}

\begin{figure}
\includegraphics[width = \linewidth]{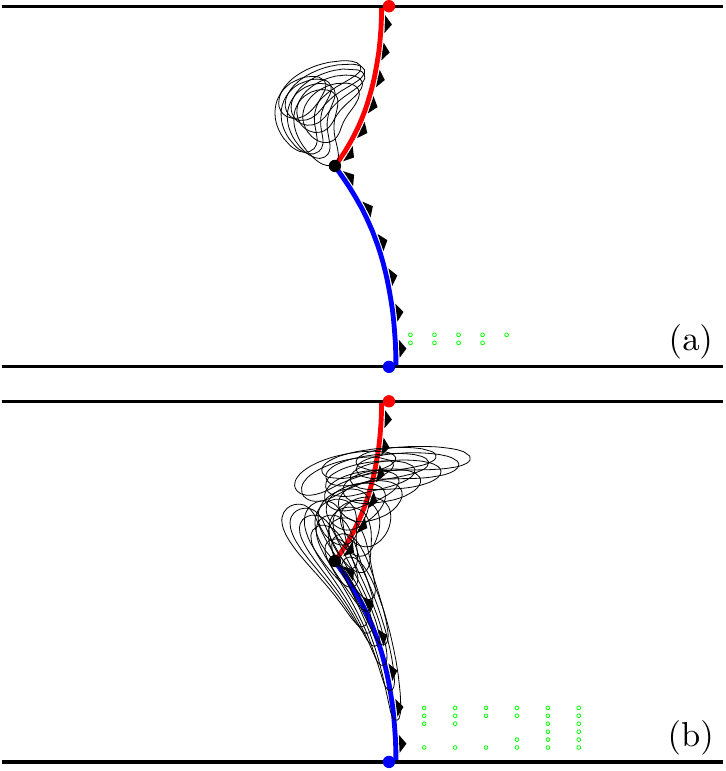}
\caption{Illustration of capture lobe function. (a) Each stimulation within $C_{-1}$ evolves entirely into the $LHS$. (b) Stimulations not within $C_{-1}$ maintain some intersection with the $RHS$. $(v_0=0.05, b=0.3, \omega=5, \phi=\pi)$}
\label{fig:capture}
\end{figure}

The capture process is slightly subtler due to the asymmetry in the burning process (compare Eqs.~(\ref{eqn:passive_lobe_defs_pts})),
%
\begin{equation}
\begin{aligned}
\label{eqn:active_lobe_defs_ptsC}
C_{-1} &= \{p \in RHS : \Mapii(p) \cap RHS = \emptyset\},\\
C_{0} &= \{p \in LHS : \forall p' \in LHS , p \not \in \Mapii(p')\}.
\end{aligned}
\end{equation}
The capture lobe $C_{-1}$ is the set of all points right of the boundary such that the evolution of this point stimulation under $\Masii$ lies completely on the left. In other words, $C_{-1}$ contains all those stimulations on the right that do not contribute to the subsequent burning of the $RHS$.
The capture lobe $C_{0}$ is the set of all points on the left not reachable in one period by any stimulation on the left, i.e. reachable only by a stimulation on the right.
See Fig.~\ref{fig:capture}.

We reiterate that, just as in the passive case, these set-based definitions of lobes are well-defined with respect to an arbitrary transport boundary. However, we again wish to make use of the dynamic structure of the system in choosing the transport boundary in a natural way.


\subsection{Recast burning lobe definitions}
The burning lobe definitions are phrased in terms of mapping properties of point stimulations.
Experimentally, this definition makes good sense because a point's lobe membership is based on the results of an elementary experiment---the stimulation of just this point (or as close to it as possible).

However, making use of the extension of $\Mapii$ to regions, we can rephrase the definition of the burning lobes. The first two, $E_0$ and $C_0$ are relatively intuitive,
\begin{equation}
\begin{aligned}
E_{0} &= RHS \cap \Masii(LHS),\\
C_{0} &= LHS \cap \Masii(LHS)^c,
\end{aligned}
\label{eqn:active_lobe_defs_extendedE0C0}
\end{equation}
where $A^c$ indicates the set complement of $A$.

\subsubsection{Two ``burning'' inverses}

In order to similarly describe the other two lobes, we need a map inverse. Since $\Masii$ is many-to-one, we must \emph{choose} an inverse. 

Given a final burned region $B$ there are an infinite number of initial regions $A$ for which $\Masii(A) = B$. Two useful senses of inverse are derived from considering extreme cases. We define the \emph{loose} and \emph{tight} inverses respectively as,
\begin{equation}
\begin{aligned}
\Masiiloose(A) &\equiv \{p : \Mapii(p) \cap A \neq \emptyset \},\\
\Masiitight(A) &\equiv \{p : \Mapii(p) \subset A \}.
\end{aligned}
\end{equation}
The loose preimage is the set of all stimulations that intersect the target, while the tight preimage is the set of only those stimulations that map entirely into the target. It follows that the tight preimage is a subset of the loose one, $\Masiitight(A) \subset \Masiiloose(A)$. (This motivates the overline versus underline notation.) Given any $A$, the loose preimage always has non-zero area while the tight preimage may be empty.
Note that $\Masiiloose(A)$ and $\Masiitight(A^c)$ form a disjoint partition of the entire space.

We may now write,
\begin{equation}
\begin{aligned}
E_{-1} &= \Masiiloose(RHS) \cap LHS,\\
C_{-1} &= \Masiitight(LHS) \cap RHS.
\end{aligned}
\label{eqn:active_lobe_defs_extendedEn1Cn1}
\end{equation}

\section{BIM basics}
\label{sec:BIM_basics}
\begin{figure}
\includegraphics[width=\linewidth]{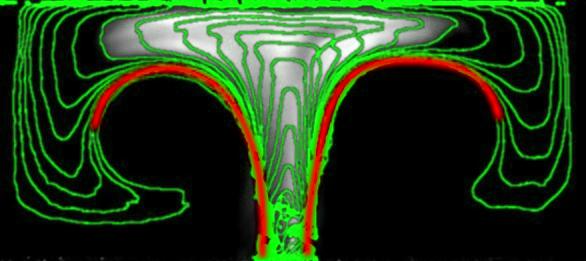}
\caption{A chemical reaction front is stimulated at the bottom of the time-independent AVC near a hyperbolic fixed point. The reacted region is advocated upward while growing in size. Edge detection applied to this image sequence results in the green contours. We deduce the presence of BIMs (red) bounding the front progress. (Reprinted with permission from Mahoney et al., EPL, 98, 44005 (2012) \cite{Mahoney12}.)} 
\label{fig:time_indep_conts_w_bims}
\end{figure}

We now review the necessary features of burning invariant manifolds \cite{Mahoney12, Mitchell12b}.
While incompressible flows have only hyperbolic (SU) and elliptic (SS) fixed points, all fixed-point stability types (SSS, SSU, SUU, UUU) are possible in front element dynamics (Eq.~(\ref{eqn:3DODE})).
For the purposes of this paper, we will concern ourselves only with the SUU and SSU stability types, and only with invariant manifolds of their singleton dimensions. We will refer to these one-dimensional objects as stable and unstable \emph{burning invariant manifolds} (BIMs), respectively.
Previous theoretical and experimental studies have demonstrated that unstable BIMs are one-sided barriers to front propagation in fluid flows, similar to traditional advective invariant manifolds (Fig.~\ref{fig:time_indep_conts_w_bims}).

While in many ways BIMs function like their passive counterparts, there are some important differences.
While advective invariant manifolds are codimension 1, BIMs are codimension 2 since they exist a three-dimensional phase space.
(Note that the figures in this work portray \emph{projections} of BIMs into $xy$-space.) 
It is then not obvious how BIMs can function as barriers.
The answer is that only a subset of curves $(x(\lambda), y(\lambda), \theta(\lambda))$ correspond to fronts. (Here $\lambda$ is the $xy$ euclidean length along the curve.)
The \emph{front-compatibility criterion} requires that $\theta(\lambda)$ everywhere equals the angle $\theta'$ describing the local tangent to the curve's $xy$-projection, $\tan \theta'(\lambda) = (d y / d \lambda) / (d x / d \lambda)$.
Any parameterized curve in $xy$-space, including the boundary curve of a burned region, uniquely lifts to a curve in $xy\theta$-space that satisfies the front-compatibility-criterion.
We shall use the term \emph{front} for both the curve in $xy$-space and its lift to $xy\theta$-space.

It has been shown that BIMs satisfy the front-compatibility criterion and so are (locally) fronts \footnote{We say \emph{locally} because unlike the curve in 3D corresponding to the perimeter of a compact burned region, a BIM cannot form a topological circle.}.
As such, BIMs are also subject to the \emph{no-passing lemma}---no front can overtake another front from behind. When the leading front is an invariant one (a BIM), the other is bounded by it. Because unstable BIMs are transversely stable, not only do they bound fronts, but fronts also converge to them making unstable BIMs evident in observations of the dynamics. A similar effect is observed in passive flows \cite{Rom-Kedar90c}.

Just as fronts have a burning direction, so do BIMs. A front near a BIM with the same burning direction will not pass through. Conversely, a front with burning direction opposite a BIM can pass through with no singularity in its motion. This is why BIMs are referred to as \emph{one-sided} barriers.

In the full $xy\theta$-space, BIM are smooth, non-intersecting curves.
However, when projected into the $xy$-plane, BIMs have cusps and self-intersections. These cusps are physically meaningful in that fronts can wrap around them.
Therefore a BIM loses its bounding behavior at a cusp.

Having established that BIMs are barriers to front propagation, just as advective invariant manifolds are barriers to fluid transport, it might seem that the analogy is complete and all that remains is to construct an active transport boundary from BIM segments. However, in the next few sections we illustrate several new features of the active system that require consideration before the analogy is made complete. We proceed by first analyzing the $v_0<<1$ regime.

\section{Small $v_0$: no new intersections}
\label{sec:small_v_0}

Consider finite segments of a generic advective invariant manifold tangle. Then there exists a $v_0$ small enough such that  taking the advective manifolds into the left- and right-burning BIMs by a continuous deformation generates no new intersections in these finite segments (see Fig.~\ref{fig:advective_turnstile_split}). Examining this simplest topological case reveals two important features in the composition of the active transport boundary.

\subsection{Co-orientation}
\begin{figure}
\includegraphics[width=\linewidth]{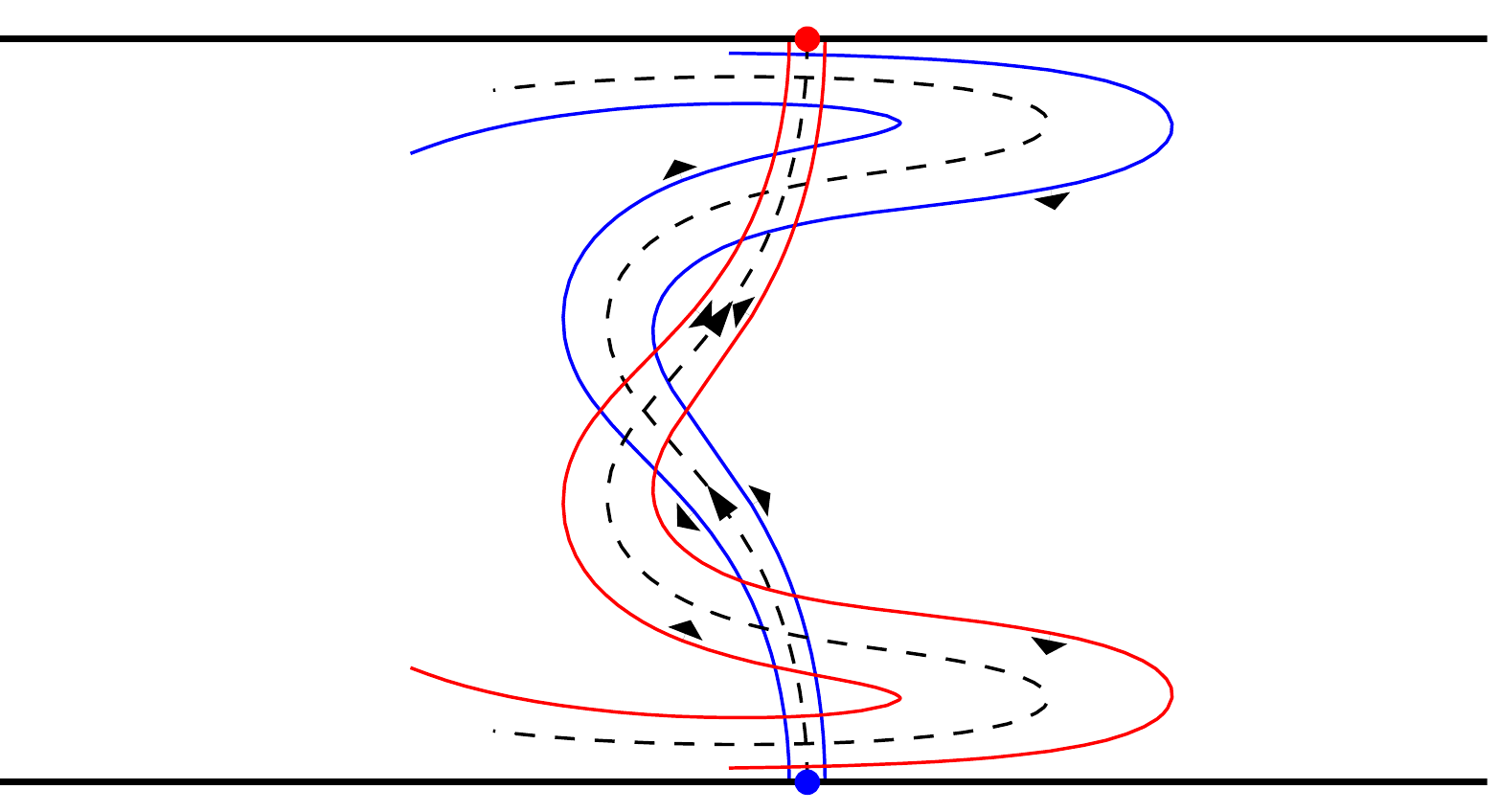}
\caption{Each advective manifold (dashed) splits into two oppositely burning BIMs (direction denoted by wide arrows). Stable (unstable) manifolds are attached to the upper (lower) wall. Note the unstable BIMs burn away from the advective manifold while the stable BIMs burn toward it. For small $v_0$, BIMs follow near the advective manifolds for a long length.  $(v_0=0.05, b=0.4, \omega=5, \phi=\pi)$}
\label{fig:advective_turnstile_split}
\end{figure}

\begin{figure}
\includegraphics[width=\linewidth]{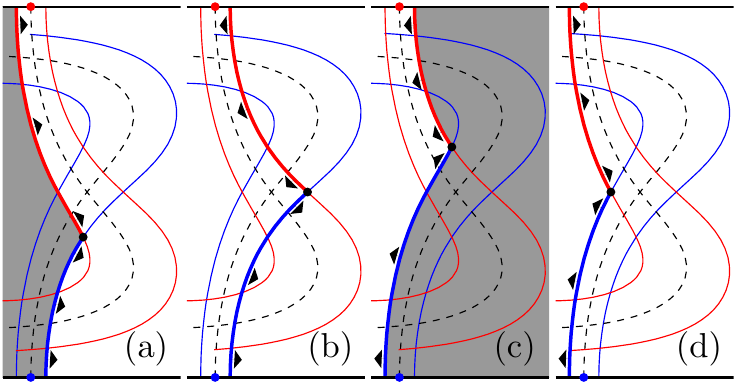}
\caption{For cases (a,c), the two BIMs agree on a burning direction and so are \emph{co-oriented}. In cases (b,d), the two BIMs are mismatched. $(v_0=0.1, b=0.3, \omega=5, \phi=0)$}
\label{fig:coorientation}
\end{figure}

As $v_0$ increases from $0$, each stable and unstable advective invariant manifold splits into nearby left- and right-burning BIMs (Fig.~\ref{fig:advective_turnstile_split}). In joining segments of these BIMs to form a transport boundary, each offers a choice of left- or right-burning.
Figure~\ref{fig:coorientation} illustrates these four choices. In case (a), the BIM segments are both rightward burning, corresponding to a burned region impinging from the left. Case (c) corresponds to the boundary of a region on the right. Because in each of these two cases, the two BIM segments agree on a burning direction, we call these configurations of joined segments \emph{co-oriented}. In cases (b) and (d), the BIM segments chosen are not co-oriented, and do not correspond to the oriented boundary of a burned $LHS$ or $RHS$. 

More precisely, consider the two front elements involved in the intersection and their respective directions of propagation $\hat{\mathbf{n}}_i$. For each, denote by $\hat{\mathbf{t}}_i$ the direction tangent to the corresponding BIM segment and oriented away from the intersection. If the two cross-products $\hat{\mathbf{n}}_i \times \hat{\mathbf{t}}_i$ are of opposite sign, the segments are co-oriented.

If we were to choose a non-co-oriented configuration (say case (b)), then a burned region abutting this curve from the left (right) will burn through the stable (unstable) BIM segment immediately. Consequently, neither this BIM segment, nor its iterate would bound the iterated burned region.
Therefore, our first new requirement of the active transport boundary is that it be composed of BIM segments that are co-oriented.
For the remainder of this paper we analyze fronts progressing from left to right as in Fig.~\ref{fig:coorientation}(a).

\subsection{Intersections: splitting, continuation, connection, concavity, convexity}

In the advective turnstile, an intersection between stable and unstable invariant manifolds must map to another such intersection. This fact allows us to neatly equate the boundary segments of a lobe iterate with the iterates of the initial boundary segments.

\begin{figure}
\includegraphics[width = \linewidth]{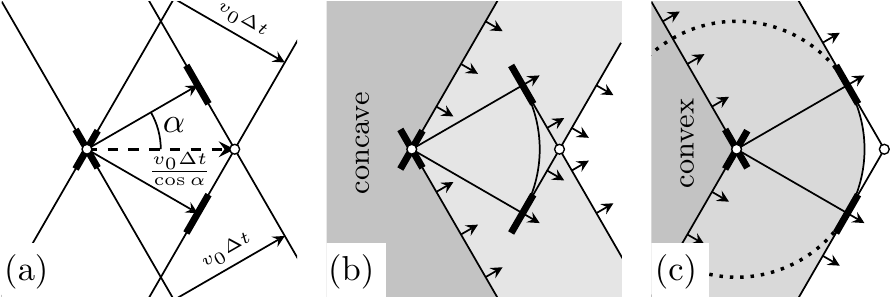}
\caption{(a) Two fronts intersect. As they propagate forward, their intersection moves faster than $v_0$. The initially intersecting front elements are now separated. (b) Concave corner. The front elements initially at the intersection become enveloped in the reacted region. The intersection continuation remains physically relevant. (c) Convex corner. Front elements at intersection remain on the physical boundary. The intersection continuation runs ahead of the reacted region leaving an unburned swallowtail shape behind. The arc is not the evolution of either front segment.}
\label{fig:scissor_and_con}
\end{figure}

In the active system, an intersection in the $xy$-projection of two BIMs is not an intersection in $xy\theta$-space. Generically, the BIM projections meet with different orientations and so two \emph{different} front elements participate in their $xy$-intersection. More importantly, these two front elements evolve forward splitting apart in $xy$-space (Fig.~\ref{fig:scissor_and_con}(a)). This might seem to foil an active generalization of the turnstile. However, the bounding property of both advective and active invariant manifolds is not a pointwise property, but a property of curve segments.
This implies that the functional intersection is not necessarily lost; we can define an \emph{intersection continuation} by tracking the continuous evolution of the intersection of the two BIMs.
Figure~\ref{fig:scissor_and_con}(a) shows that while the two front elements initially involved in the intersection always separate, a natural continuation of this intersection can be defined. This continuation outruns each of the two front elements by a factor of $1/\cos \alpha$ in the comoving fluid frame.

Figure~\ref{fig:scissor_and_con}(a) is compatible with two different burned regions. In Fig.~\ref{fig:scissor_and_con}(b), the burned region has a \emph{concave corner} while the region in Fig.~\ref{fig:scissor_and_con}(c) has a \emph{convex corner} (compare Fig.~\ref{fig:coorientation}(a,c)). In order to understand the evolution of these burned regions using the 3D ODE for front propagation, we need to properly join the two original front segments at the corner. Remembering that fronts are curves in $xy\theta$-space, it is clear that the two front segments alone form a discontinuous curve.
We join these two front segments with a \emph{connection} $\mathcal{C}$ that is simply a straight line
connecting the $\theta$ values of the two front elements at the corner.
Since $\theta \in S^1$, there are two distinct intervals between any two $\theta$ values; the physically correct choice is the shorter of the two intervals.
The two segments joined with the connection are now a continuous front in $xy\theta$-space.

In Fig.~\ref{fig:scissor_and_con}(b), the connection evolves into the arc of a circle and is immediately burned past by both front segments. Further, the front segments partially overtake each other sending finite portions into the reacted region. The boundary of the final reacted region $\Masii(R)$ is a subset of the evolved front, $\partial \Masii(R) \subsetneq \Maciii(\partial R)$.

In Fig.~\ref{fig:scissor_and_con}(c), the connection unfolds in the same way, but it remains on the boundary of the reacted region, $\partial \Masii(R) = \Maciii(\partial R)$. This connection, functioning like a point stimulation made at the convex corner, is a boundary between burned and unburned that is not the iterate of, or continuation of, either original front segment; it generates ``new boundary''. Also notice that the intersection continuation outruns this point stimulation. This concept will lead to the second requirement of the active transport boundary.

\subsection{pips: splitting and continuation}
Having now described front element splitting and intersection continuation and the effect of concavity on the physical relevance of the connection, we apply this to pips between the BIMs. First we establish some notation for pips, comparing the passive and active case.

Recall that for advective dynamics, a  point $p$ is an \emph{advective pip} if $W^S[z^A, p] \cap W^U[z^B, p] \ \backslash \ \{z^A, z^B, p\}=\emptyset$ (see Fig.~\ref{fig:advective_turnstile}).
The image and preimage of a advective pip $p_0$ are denoted $\Mpp(p_0) = p_1$ and $\Mppinv(p_0) = p_{-1}$, respectively. For any advective pip $p_k$, $\Mpp^n(p_k) = p_{k+n}, n,k \in \mathbb{Z}$.

\begin{figure}
\includegraphics[width=\linewidth]{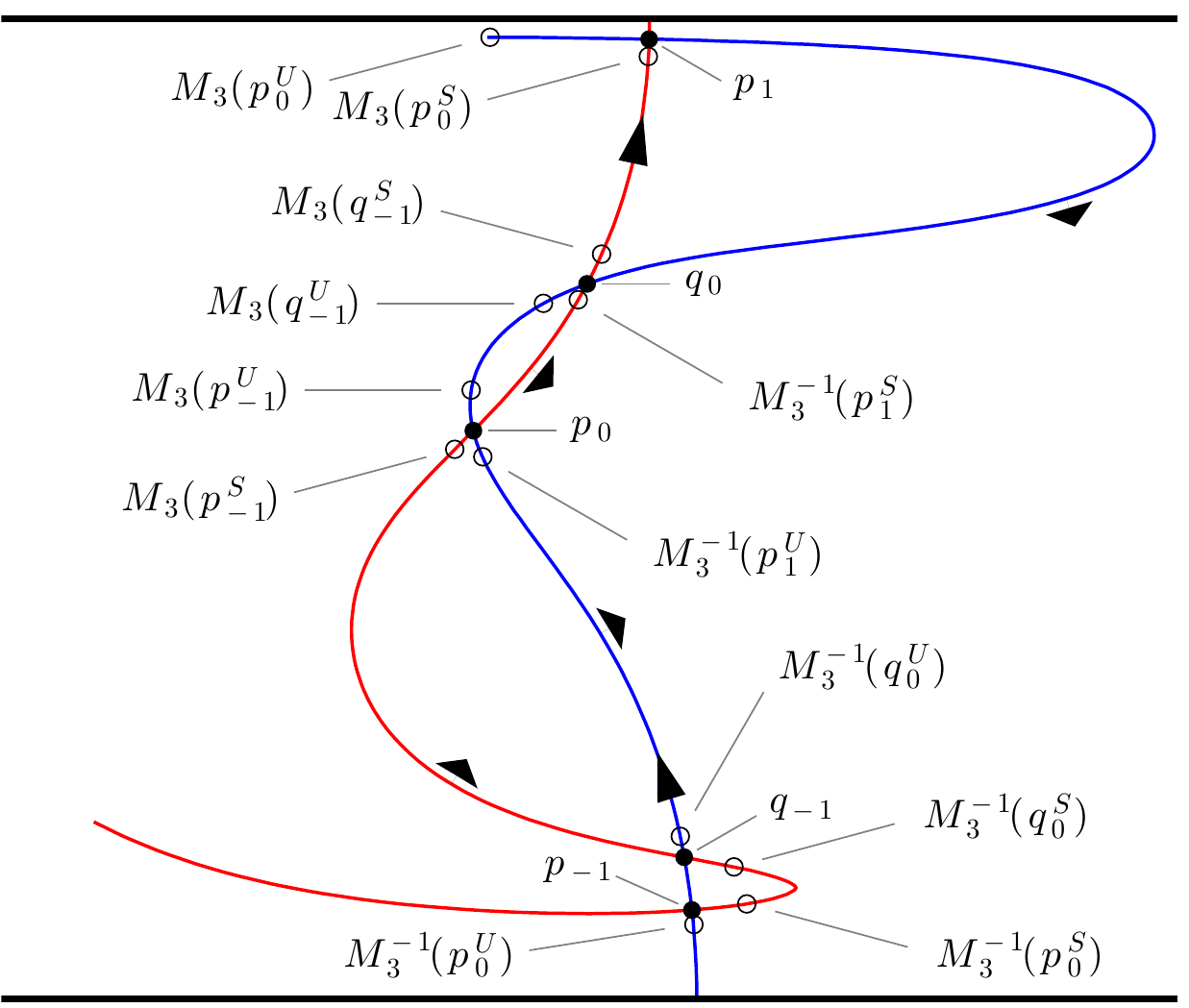}
\caption{Demonstration of the splitting of intersection points. Intersection continuations $p_i$ and $q_i$ (solid dots) represent two coincident front elements. These elements are mapped forward and backward (open dots). $(v_0=0.05, b=0.4, \omega=5, \phi=\pi)$}
\label{fig:ps_and_qs}
\end{figure}

We now consider pips in the active case. The tuple of two intersecting front elements $p \equiv (p^S, p^U)$, where $p^S, p^U \in \mathbb{R}^2 \times S^1$, is a \emph{burning pip} if $\Pi_{xy} (W^S[z^A, p^S]) \cap \Pi_{xy}(W^U[z^B, p^U]) \ \backslash \ \Pi_{xy}(\{z^A, z^B, p\})=\emptyset$ where $\Pi_{xy}$ denotes projection into the $xy$-plane.
In Fig.~\ref{fig:ps_and_qs} we have labeled the burning pips $p_i$ and $q_i$ (solid dots).
We write $p^S_i$ ($p^U_i$) to indicate the stable (unstable) component of the $i$th burning pip $p_i \equiv (p^S_i, p^U_i)$, and similarly for $q_i$.
Since a burning pip $p_0$ splits, $\Mapiii(p^S_0) \neq p^S_1$, $\Mapiii(p^U_0) \neq p^U_1$, $\Mapiiiinv(p^S_0) \neq p^S_{-1}$, and $\Mapiiiinv(p^U_0) \neq p^U_{-1}$. 
Generically, $\Mapiii^n(p_k) \neq p_{k+n}, \forall n,k \in \mathbb{Z}, n \neq 0$. 
However, continuation of intersections takes $p_i \to p_{i+1}$ and $q_i \to q_{i+1}$.
In Fig.~\ref{fig:ps_and_qs} unfilled dots denote iterates of the two components of each burning pip (solid dots) along either the stable or unstable BIM.

For instance, examine the evolution of $p_{-1}$ in Fig.~\ref{fig:ps_and_qs}. Because of front element splitting, $\Mapiii(p_{-1}) \neq p_{0}$. Instead, the two front elements $p^S_{-1}$ and $p^U_{-1}$ that compose $p_{-1}$ map to $\Mapiii(p^S_{-1})$ and $\Mapiii(p^U_{-1})$. Notice that the two points $\Mapiii(p^S_{-1})$ and $\Mapiii(p^U_{-1})$ have mapped, respectively, behind the local unstable and stable BIMs at $p_0$. 

Unlike the passive case, in which the boundary of a lobe exactly maps to the boundary of the iterated lobe, in the active case, the stable (unstable) boundary segment of a lobe maps to a stable (unstable) segment that is either shorter or longer than the stable (unstable) boundary of the subsequent lobe.
We use this knowledge, and the fact that pips alternate between concave and convex, to derive the following relations. 
\begin{equation}
\begin{aligned}
\Maciii(W^S[p_{-1},q_{-1}]) &= W^S[\Mapiii(p^S_{-1}), \Mapiii(q^S_{-1})]\\
&\supsetneq W^S[p_{0},q_{0}]\\
\Maciii(W^U[p_{-1},q_{-1}]) &= W^U[\Mapiii(p^U_{-1}), \Mapiii(q^U_{-1})]\\
&\subsetneq W^U[p_{0},q_{0}]\\
\Maciii(W^S[q_{-1},p_{0}]) &= W^S[\Mapiii(q^S_{-1}), \Mapiii(p^S_{0})]\\
&\subsetneq W^S[q_{0},p_{1}]\\
\Maciii(W^U[q_{-1},p_{0}]) &= W^U[\Mapiii(q^U_{-1}), \Mapiii(p^U_{0})]\\
&\supsetneq W^U[q_{0},p_{1}]
\end{aligned}
\end{equation}

\subsection{Boundaries of the lobes $E_0$ and $C_0$}

\subsubsection{concave pip ensures BIM boundary}

We now show that constructing a transport boundary from BIM segments joined by a concave burning pip ensures that the $E_0$ and $C_0$ burning lobes are bounded by BIM segments. In reference to Fig.~\ref{fig:ps_and_qs}, we construct the transport boundary from $W^S[z^A, p_0]$ and $W^U[z^B, p_0]$ and the concave connection $\mathcal{C}$ (Fig.~\ref{fig:scissor_and_con}(b)).

\begin{figure}
\includegraphics[width=\linewidth]{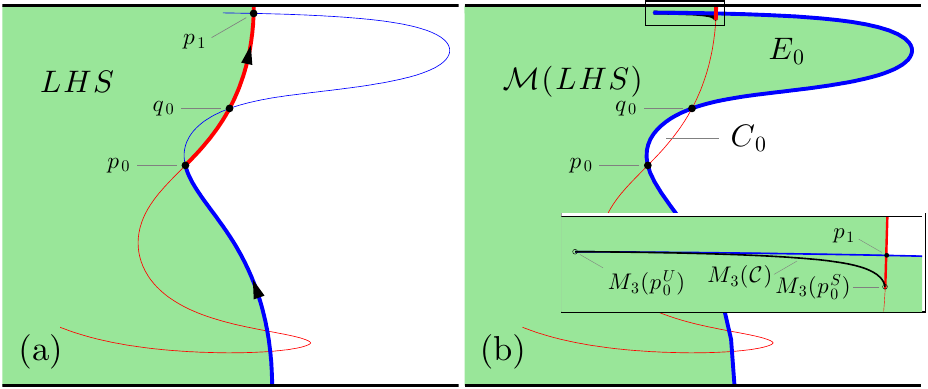}
\caption{(a) A concave pip is chosen to define the transport boundary. (b) $LHS$ and $TB$ are evolved forward showing the $E_0$ and $C_0$ lobes. $(v_0=0.05, b=0.4, \omega=5, \phi=\pi)$}
\label{fig:LHS_forward}
\end{figure}

This transport boundary ($TB$) is the bounding front of $LHS$, and so by evolving $TB$ forward, we can understand the boundary of the evolution of $LHS$. Because we chose a concave pip, $\partial \Masii(LHS) \subsetneq \Maciii(\partial LHS)$. We know by the invariance of BIMs and Fig.~\ref{fig:ps_and_qs} that this transport boundary will map forward to $W^S[z^A, \Mapiii(p^S_0)] \cup \Maciii(\mathcal{C}) \cup W^U[z^B, \Mapiii(p^U_0)]$. We also know that the subsection of this front $W^S[p_1, \Mapiii(p^S_0)] \cup \Maciii(\mathcal{C}) \cup W^U[p_1, \Mapiii(p^U_0)]$ will lie behind the boundary of the burned region $W^S[z^A, p_1] \cup W^U[z^B, p_1]$ (see Fig.~\ref{fig:LHS_forward}).

Recalling the set definition of the $E_0$ lobe (Eq.~(\ref{eqn:active_lobe_defs_extendedE0C0})), we see in Fig.~\ref{fig:LHS_forward} that $E_0 = \Masii(LHS) \cap RHS$ is bounded by $W^S[q_0, p_1]$ and $W^U[q_0, p_1]$. 
It is similarly straightforward to see that $C_0 = \Masii(LHS)^c \cap LHS$ is bounded by $W^S[p_0, q_0]$ and $W^U[p_0, q_0]$. 

\subsubsection{convex pip creates non-BIM boundary segment}

\begin{figure}
\includegraphics[width=\linewidth]{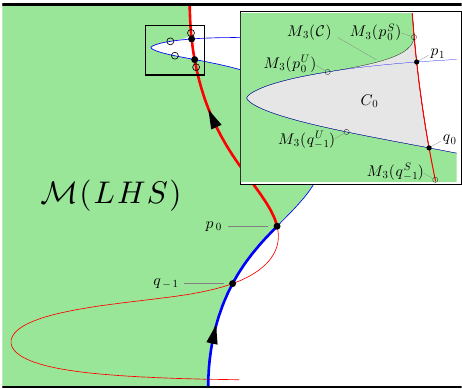}
\caption{Result of choosing convex pip for transport boundary (bold). Start with $LHS$ burned, and iterate (filled green). Small unburned region (zoomed inset) has one boundary arc which is not a BIM.  $(v_0=0.05, b=0.4, \omega=5, \phi=0)$}
\label{fig:wrong_pip_bad_C0}
\end{figure}

If a convex pip is chosen instead ($p_0$ in Fig.~\ref{fig:wrong_pip_bad_C0}), the convex connection $\mathcal{C}$ used to join $W^S[z^A, p_0]$ and $W^U[z^B, p_0]$ leads to the following problem.
The connection evolves into the arc $\Maciii(\mathcal{C})$ (Fig.~\ref{fig:wrong_pip_bad_C0} inset), which lags behind the continuation $p_1$.
The space between $\Maciii(\mathcal{C})$ and the local stable and unstable segments near $p_1$ is part of $LHS$ yet unburned, and therefore part of the $C_0$ lobe.
Thus the boundary of $C_0$ includes the arc $\Maciii(\mathcal{C})$, which is not a BIM segment.
Therefore, in order to maintain the property that the $E_0$ and $C_0$ lobe boundaries are BIM segments, we must impose the second constraint that the transport boundary is defined by a \emph{concave} burning pip.

\subsection{Relations between fronts of preimages and preimages of fronts}

\begin{figure}
\includegraphics[width=\linewidth]{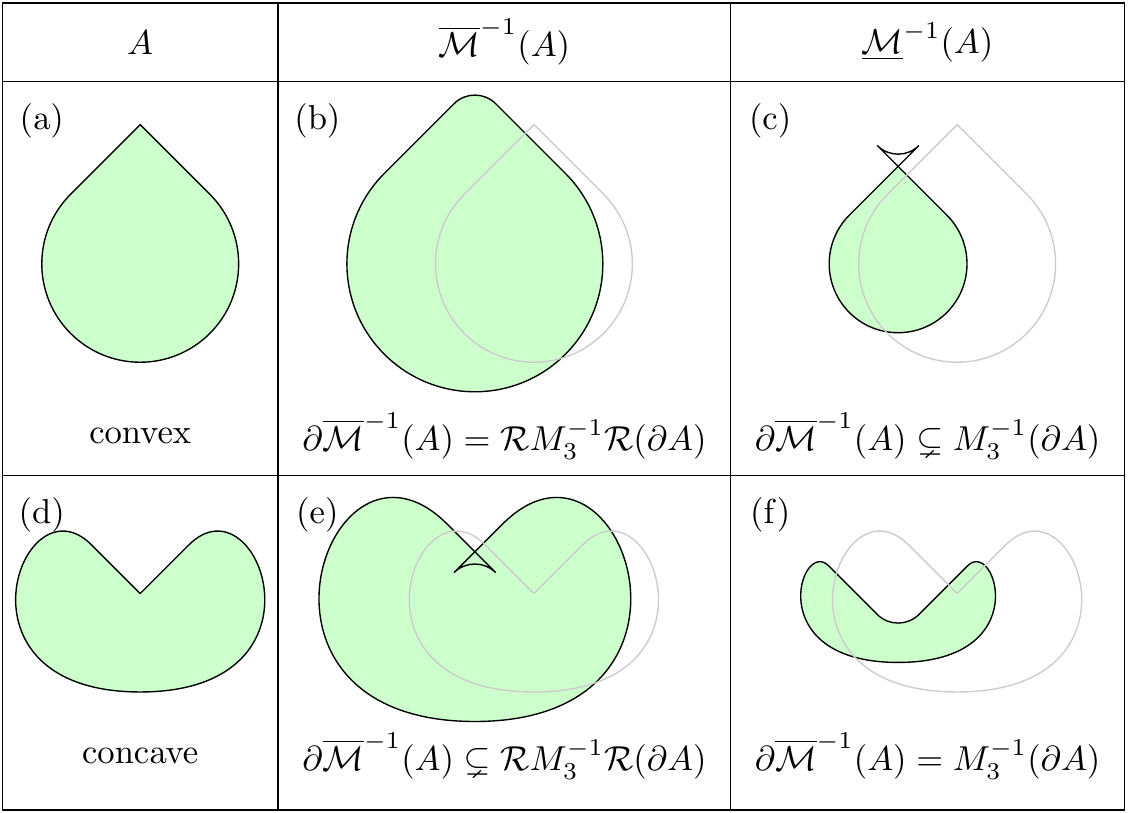}
\caption{Illustration of how convex (a) and concave (d) corners behave under loose (b,e) and tight (c,f) preimages. $\mathbf{u}$ is a uniform flow to the right. (b,c,e,f) Green regions are preimages of $A$. 
Black curves represent two ways of mapping $\partial A$ backwards. Reference curves (gray) show original region.}
\label{fig:preimages}
\end{figure}

The correspondence between burned region evolution (Eq.~(\ref{eqn:FKPP})) and bounding front propagation (Eq.~(\ref{eqn:3DODE})) has already been established. We used this correspondence to understand the nature of the $E_0$ and $C_0$ lobes. Here we establish an analogous correspondence between the loose and tight preimages of a region $A$ and the bounding curves so that we may similarly understand the $E_{-1}$ and $C_{-1}$ lobes.
Figure~\ref{fig:preimages} illustrates sets $A$, $\Masiiloose(A)$ and $\Masiitight(A)$ as well as the relevant curves.

We begin with the loose preimage. Consider the set of stimulations which do not reach the target set $A$, $D = \{p: \Masii(p) \cap A = \emptyset\} = \Masiiloose(A)^c$. 
The boundary of $D$ maps into the boundary of $A$ with orientation reversed, $\Maciii(\partial D) \subset \Maciiirevburn(\partial A)$. Application of an inverse and second reversal yields $\Maciiirevburn(\partial D) \subset \Maciiirevburn \Maciiiinv \Maciiirevburn(\partial A)$. Finally, since $D$ is the complement of the set of interest, $\Maciiirevburn(\partial D) = \partial \Masiiloose(A)$. Thus we have the following relation between the boundary of the loose preimage and $A$ through the front element dynamics,
\begin{align}
\partial \Masiiloose(A) \subset \Maciiirevburn \Maciiiinv \Maciiirevburn(\partial A).
\label{eqn:loose_boundary_relation}
\end{align}
Equality is obtained when no local (swallowtails) or global front intersections develop in $\Maciiirevburn \Maciiiinv \Maciiirevburn(\partial A)$ (equivalently in $\Maciiiinv \Maciiirevburn(\partial A)$). As we see comparing Figs.~\ref{fig:preimages}(b) and \ref{fig:preimages}(e), a concave corner is certain to produce a swallowtail and break the equality.

The tight preimage relation is simpler to state,
\begin{align}
\partial \Masiitight(A) \subset \Maciiiinv (\partial A),
\label{eqn:tight_boundary_relation}
\end{align}
similarly with equality if no front intersections are developed in $\Maciiiinv (\partial A)$.


\subsection{Boundaries of the lobes $E_{-1}$ and $C_{-1}$}

\begin{figure}
\includegraphics[width=\linewidth]{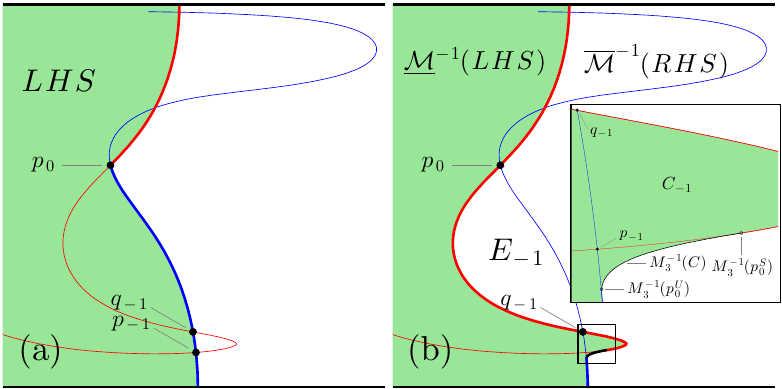}
\caption{(a) A concave pip is chosen to define the transport boundary. (b) $RHS$ and $TB$ are evolved backward (loose preimage) showing the $E_{-1}$ and $C_{-1}$ lobes.
$E_{-1}$ lobe is bounded by BIMs. $C_{-1}$ lobe has non-BIM boundary. $(v_0=0.05, b=0.4, \omega=5, \phi=\pi)$}
\label{fig:LHS_backward}
\end{figure}

With the correspondence between bounding fronts and region preimages (Fig.~(\ref{fig:preimages})) established, we can determine the $E_{-1}$ and $C_{-1}$ lobes with the assistance of the invariant manifolds. 
In Fig.~\ref{fig:LHS_backward}(a), we define the transport boundary $TB$ using a concave pip.

The $E_{-1}$ lobe depends on the loose preimage $\Masiiloose(RHS)$ through Eq.~(\ref{eqn:active_lobe_defs_extendedEn1Cn1}).
Let us examine this preimage.
In the previous section, we established the relation Eq.~(\ref{eqn:loose_boundary_relation}) between the boundary of a preimage and inverse images of boundaries. Here this relation gives,
\begin{align*}
\partial \Masiiloose(RHS) &\subset \Maciiirevburn \Maciiiinv \Maciiirevburn(\partial RHS)\\
&= \Maciiirevburn \Maciiiinv (TB)
\end{align*}
Figure~\ref{fig:LHS_backward}(b) illustrates the inverse of the transport boundary $\Maciiiinv (TB)$ and the region it bounds (green). 
Since $TB$ has a concave pip, $\Maciiiinv (TB)$ is not forced to generate a swallowtail. 
Our earlier assumption about $v_0$ being small enough to not generate any new intersections ensures that the entire curve $\Maciiiinv (TB)$ is free of self intersections.
Thus we have the equality $\partial \Masiiloose(RHS) = \Maciiirevburn \Maciiiinv (TB)$.
This ensures that the loose preimage of $RHS$ is exactly the white region in Fig.~\ref{fig:LHS_backward}(b).
Finally intersecting with $LHS$ (Eq.~(\ref{eqn:active_lobe_defs_extendedEn1Cn1})), we see the location of $E_{-1}$ and that its boundaries are the BIM segments $W^S[q_{-1}, p_0]$ and $W^U[q_{-1}, p_0]$.

On the other hand, $C_{-1}$ depends on the tight preimage of the other side, $\Masiitight(LHS)$ (Eq.~(\ref{eqn:active_lobe_defs_extendedEn1Cn1})). Equation~(\ref{eqn:tight_boundary_relation}) gives,
\begin{align*}
\partial \Masiitight(LHS) &\subset \Maciiiinv (\partial LHS)\\
&= \Maciiiinv (TB)
\end{align*}

Therefore, $\Masiitight(LHS)$ is exactly the green shaded region in Fig.~\ref{fig:LHS_backward}(b).
Intersecting with $RHS$, we see that $C_{-1}$ is not bounded by BIM segments $W^S[p_{-1}, q_{-1}]$ and $W^U[p_{-1}, q_{-1}]$ (see Fig.~\ref{fig:LHS_backward}(b) inset). It is instead bounded by $W^S[\Mapiiiinv(p^S_0), q_{-1}]$, $W^U[\Mapiiiinv(p^U_0), q_{-1}]$ and $\Maciiiinv(\mathcal{C})$. In the same way, we previously found that when a convex pip is used, the $C_0$ lobe is bounded in part by $\Maciii(\mathcal{C})$, a non-BIM segment.

If one is only interested in the escape process, i.e. transport of the 
burned region to the right of the $TB$, both concave and convex pips are 
equally suitable, in that $E_{-1}$ and $E_0$ are both bounded entirely 
by BIMs.  However, for the capture process, i.e. entrainment of unburned 
fluid to the left of the $TB$, either the lobe $C_{-1}$ (for a concave 
pip) or the lobe $C_0$ (for a convex pip) will be bounded in part by a 
non-BIM segment.  We shall preferentially consider concave pips in the 
remainder of this paper, both because the lobe $C_0$ is more directly 
visible in experiments and because the relevant $TB$ (concave) can be realized through
the dynamical development of a burned region.

\subsection{Map inclusions}
\begin{figure}
\includegraphics[width=\linewidth]{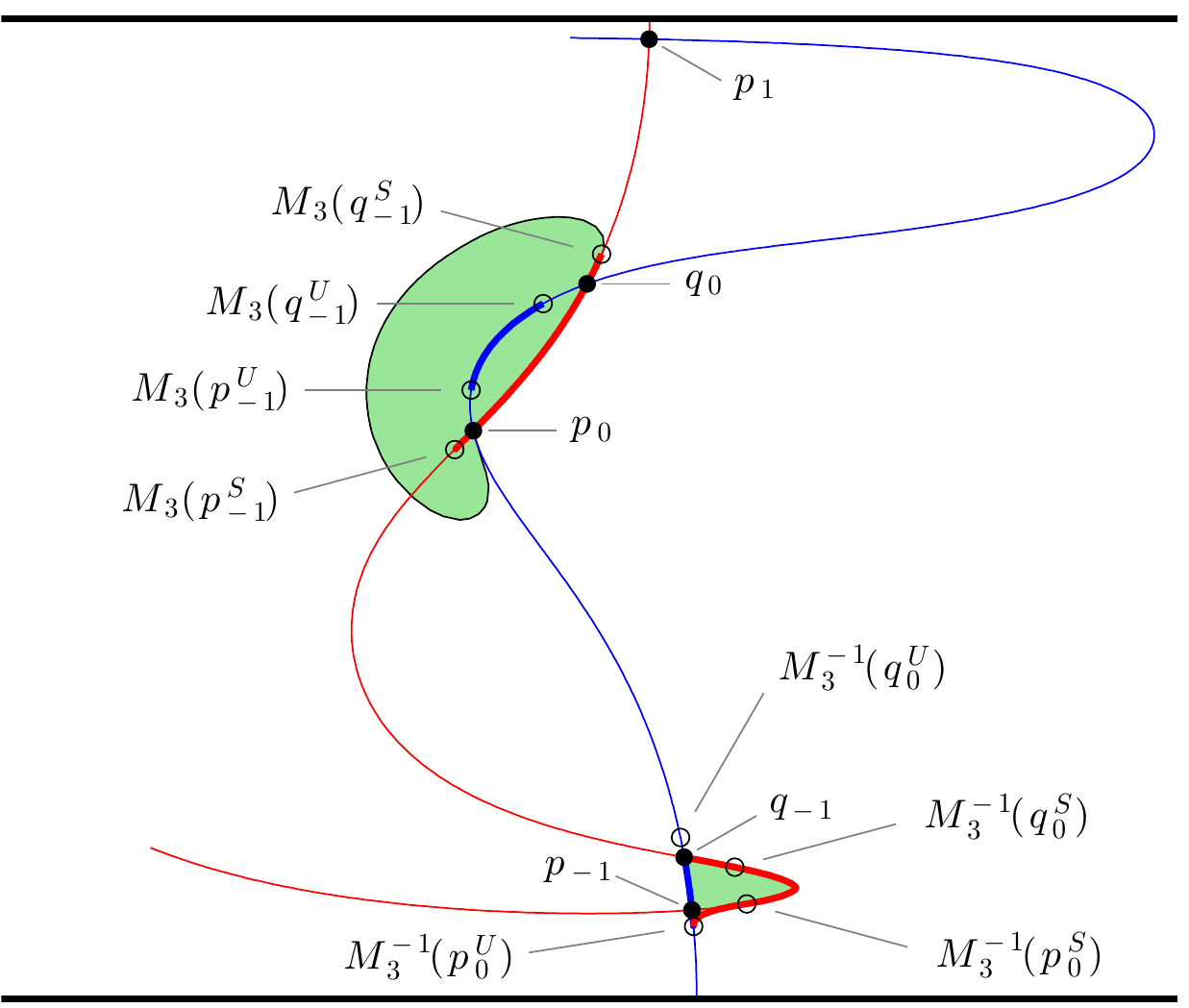}
\caption{Evolution of $C_{-1}$ lobe and its bounding segments.  $(v_0=0.05, b=0.4, \omega=5, \phi=\pi)$}
\label{fig:capture_lobe}
\end{figure}

\begin{figure}
\includegraphics[width=\linewidth]{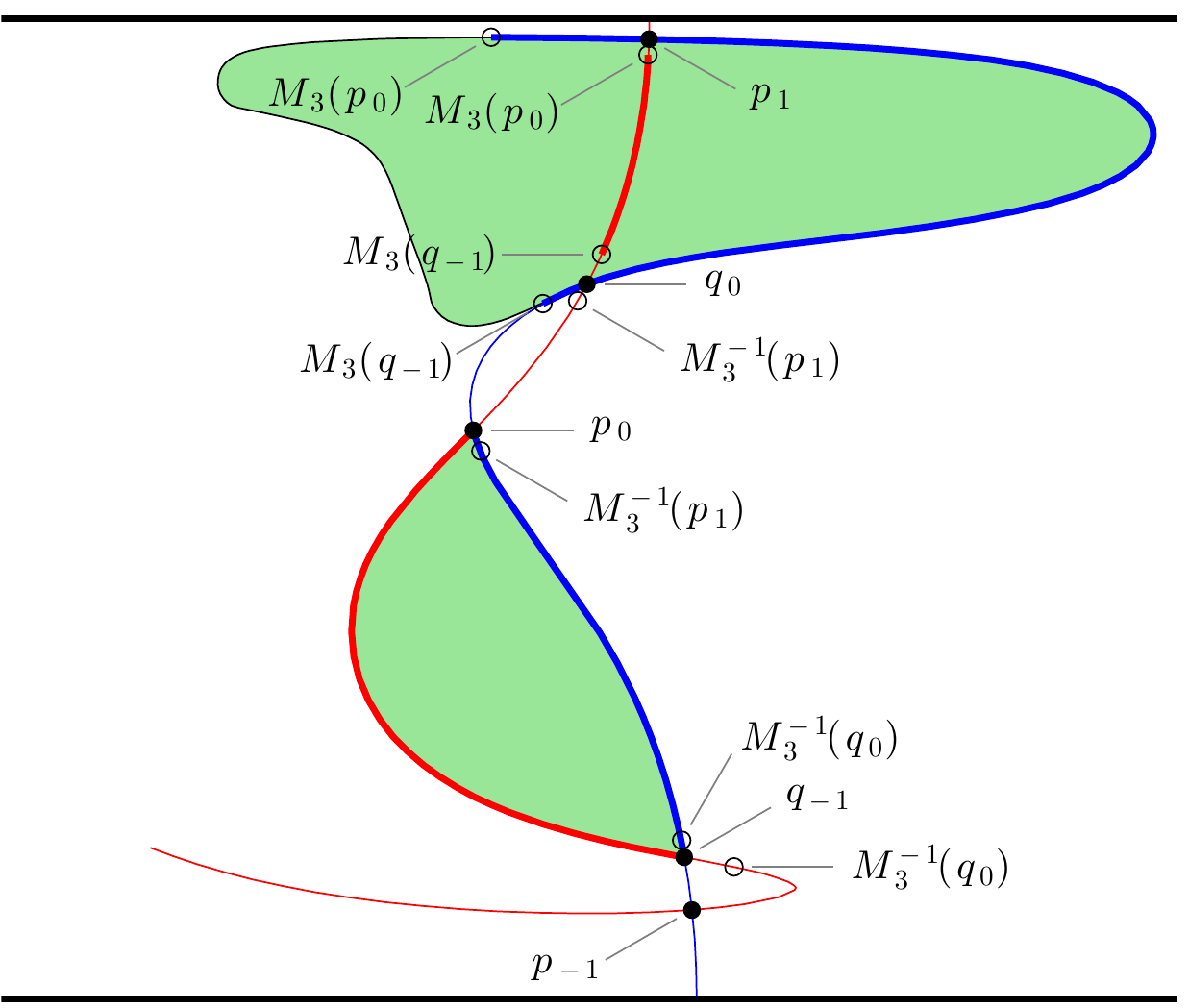}
\caption{Evolution of $E_{-1}$ lobe and its bounding segments. $(v_0=0.05, b=0.4, \omega=5, \phi=\pi)$}
\label{fig:escape_lobe}
\end{figure}

As can be seen in Figs.~\ref{fig:capture_lobe} and \ref{fig:escape_lobe}, the passive relations $C_0 = \Mps(C_{-1})$ and $E_0 = \Mps(E_{-1})$ are softened to the following inclusions,
\begin{subequations}
\label{eqn:inclusion_loose}
\begin{align}
E_0 \subset \Masii(E_{-1}), \label{eqn:inclusion_loosea}\\
C_0 \subset \Masii(C_{-1}). \label{eqn:inclusion_looseb}
\end{align}
\end{subequations}
The $E_0$ inclusion follows directly from the definition Eq.~(\ref{eqn:active_lobe_defs_ptsE}).
However, the $C_0$ inclusion depends on the fact that $\Maciii(W^S[p_{-1}^S, q_{-1}^S])$ includes $W^S[p_0, q_0]$, the stable boundary of $C_0$.
We may reobtain $E_0$ by intersecting Eq.~(\ref{eqn:inclusion_loosea}) with the $RHS$,
\begin{equation}
\begin{aligned}
E_0 = \Masii(E_{-1}) \cap RHS.
\end{aligned}
\label{eqn:inclusion_tight}
\end{equation}
%

\section{Arbitrary $v_0$: the role of additional intersections}
\label{sec:large_v_0}

We have shown that for appropriately small $v_0$, there are new criteria that need to be taken into account when constructing a burning turnstile. These criteria arise from breaking the $v_0 = 0$ symmetry.
In this section we examine the case of larger $v_0$, where new intersections arise.

\subsection{Swallowtails}

\begin{figure}
\includegraphics[width = \linewidth]{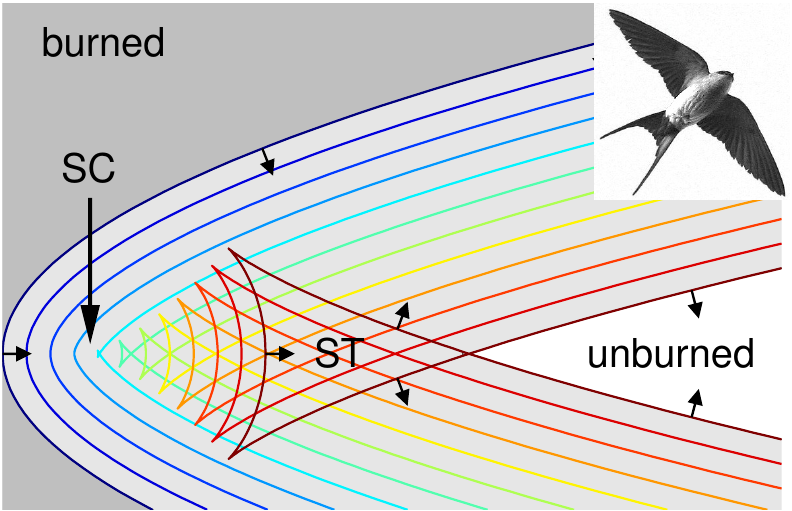}
\caption{The creation of a swallowtail in a system with no flow, as is common in standard optics and other front propagation phenomena. Burned region with parabolic boundary impinging from the left. Front elements first collide at a singularity---the swallowtail catastrophe (SC). The final swallowtail-shaped region (ST) has been traversed by two front segments. (inset: red-rumped swallow, wiki commons images)}
\label{fig:swallowtail}
\end{figure}

In classical front propagation theory, formation of swallowtail catastrophes is generic \cite{Arnold83} (Fig.~\ref{fig:swallowtail}). A front with a concave segment has a point of greatest curvature or, equivalently, minimum radius of curvature. A swallowtail catastrophe occurs when this point propagates forward to its local center of curvature. As the front propagates further still, it develops a self-intersecting structure reminiscent of a swallow's tail. In the case of a front separating burned and unburned regions, the swallowtail must lie entirely within the burned region. That is, the front segments that make up the swallowtail are no longer boundaries between burned and unburned regions.

Swallowtail formation is also generic for fronts propagating in flowing media.
The flow will influence the curvature of the front, thereby promoting or preventing formation of swallowtails.
Many flows will tend to generate and regenerate areas of concavity in propagating fronts, thus facilitating continued swallowtail formation.

\subsubsection{$C_{-1}$ lobe swallowtail}

\begin{figure}
\includegraphics[width = \linewidth]{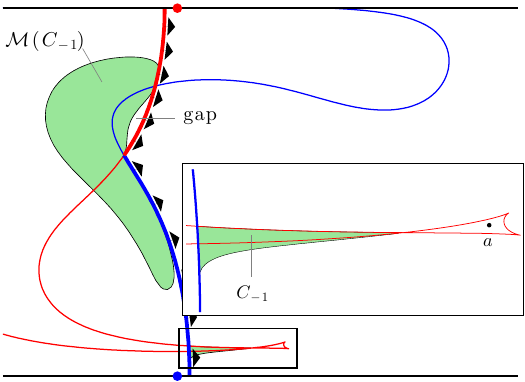}
\caption{Swallowtail diminishes size of $C_{-1}$ lobe. Evolving $C_{-1}$ forward covers most of $C_0$, but leaves an empty space along stable BIM. $(v_0=0.085, b=0.3, \omega=5, \phi=\pi)$}
\label{fig:Cn1swallowtail_w_region}
\end{figure}

Given the basic turnstile geometry in Fig.~\ref{fig:advective_turnstile}, we might expect to see swallowtails first appear in the boundary of the $C_{-1}$ lobe \footnote{It is important to note that since ray optics is invertible, reversing the propagation direction (or time) can cause the swallowtail to unfold. This phenomenon is found in the stable BIMs (see Fig.~\ref{fig:Cn1swallowtail_w_region}).}. This is because the lobe is long and narrow, creating a segment of high curvature at the tip. As the right-burning stable BIM deviates from the advective manifold (with increasing $v_0$), the BIM is forced inward, increasing the curvature at the tip until a swallowtail is forced \footnote{We are making use of the stability of this perestroika. That is, the structural change is actually occurring in time, but according to Ref.~\cite{Arnold83}, we can see the same morph with a generic parameter sweep (such as $v_0$).}.

Let us consider the effect of such a swallowtail on the $C_{-1}$ lobe. This lobe is defined as the set of all point stimulations that cross the transport boundary completely. Suppose a stimulation within the triangular region (point $a$ in Fig.~\ref{fig:Cn1swallowtail_w_region}) were to evolve completely across the $TB$. Then the last front element to pass through the stable BIM would be tangent to it and co-oriented. Hence, according to the no-passing lemma, this stimulation could not pass through entirely---a contradiction. The swallowtail region is therefore not a part of the $C_{-1}$ lobe. The area remaining is shown in Fig.~\ref{fig:Cn1swallowtail_w_region} (inset). 
The first effect of swallowtails, then, is that the $C_{-1}$ area is diminished by the encroaching swallowtail.

The second effect is that the boundary of the $C_{-1}$ lobe is reduced; here, the middle segment of the stable BIM is no longer on the lobe boundary (Fig.~\ref{fig:Cn1swallowtail_w_region} inset). In general, a single bounding BIM segment is replaced by the joining of two segments.

Evolving this reduced area forward, we find the situation in the upper-left of Fig.~\ref{fig:Cn1swallowtail_w_region}. The region bounded by stable and unstable BIMs is still the $C_0$ lobe---nothing in this region is reached by any stimulation on the $LHS$. However, notice that a small portion also remains unburned by the evolution of $C_{-1}$. We can understand this by looking at Fig.~\ref{fig:scissor_and_con}(c); the intersection continuation runs ahead of the burned region leaving an unburned gap. In this case, the intersection continuation is lost when the swallowtail itself is removed by its catastrophe; at this point, we are left with the evolution of two distinct fronts (the boundary of the evolving $C_{-1}$ lobe and the now unfolded segment of stable BIM). The arc that bounds the unburned gap in Fig.~\ref{fig:scissor_and_con}(c) is in Fig.~\ref{fig:Cn1swallowtail_w_region} geometrically distorted by the flow, reversing its concavity.

Our previous inclusion relation for the evolution of $C_{-1}$ (Eqs.~(\ref{eqn:inclusion_looseb})) has thus been weakened to: if $C_{-1} \neq \emptyset$, then $C_0 \cap \Masii(C_{-1}) \neq \emptyset$.

\subsubsection{$C_0$ lobe swallowtail}

In the same spirit, the $C_0$ lobe may form a swallowtail as the unstable BIM deviates inward from the advective unstable manifold (Fig.~\ref{fig:four_swallowtails}). This swallowtail region represents fluid that has been crossed by two separate front segments of the evolving $LHS$. As $C_0$ is the region that is not burned by any stimulation on the $LHS$, this swallowtail region is excluded from the $C_0$ lobe.

The unstable BIM segment that evolves to form this swallowtail previously bounded part of the $C_{-1}$ lobe. However note that this segment is oriented into $C_{-1}$. Therefore, evolving $C_{-1}$ forward, the whole unstable BIM segment (whether it forms a swallowtail or not) is immediately burned through by $\Masii(C_{-1})$. Therefore, this swallowtail on the \emph{unstable} manifold does not itself negate the inclusion relation $C_0 \subset \Masii(C_{-1})$.

Just as we saw in the $C_{-1}$ lobe, the existence of a $C_0$ swallowtail only removes lobe boundary and so this boundary is still composed of BIM segments.

\subsubsection{$E_{-1}$ and $E_0$ lobe swallowtails}

\begin{figure}
\includegraphics[width = \linewidth]{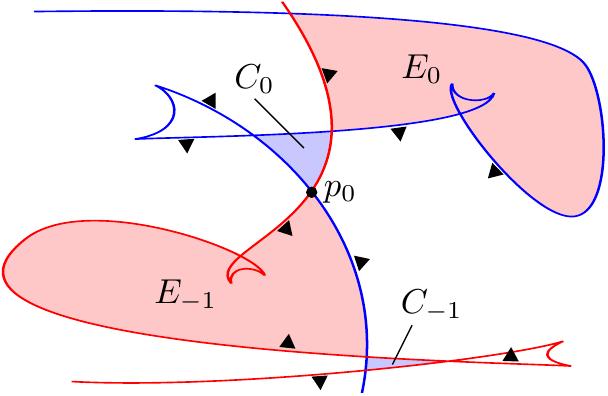}
\caption{Cartoon illustrating four basic swallowtails in the burning turnstile---one in each lobe. More complex flows can generate yet more swallowtails.}
\label{fig:four_swallowtails}
\end{figure}

Adjusting the flow appropriately, the advective $E_0$ lobe becomes longer, thinner and begins to fold. The active $E_0$ lobe can develop a swallowtail in the concave folding region (Fig.~\ref{fig:four_swallowtails}). This area is ``burned twice'' and so is still within the $E_0$ lobe. The inclusion relation remains $E_0 \subset \Masii(E_{-1})$.

The $E_{-1}$ burning lobe can be similarly modified. As in the $C_{-1}$ case, this swallowtail unfolds in forward time. 
However, since stimulations in $E_{-1}$ need only intersect the $RHS$, the swallowtail is included in $E_{-1}$.
The inclusion relation is also unaffected.
%

\subsection{Existence of pips}

\begin{figure}
\includegraphics[width = \linewidth]{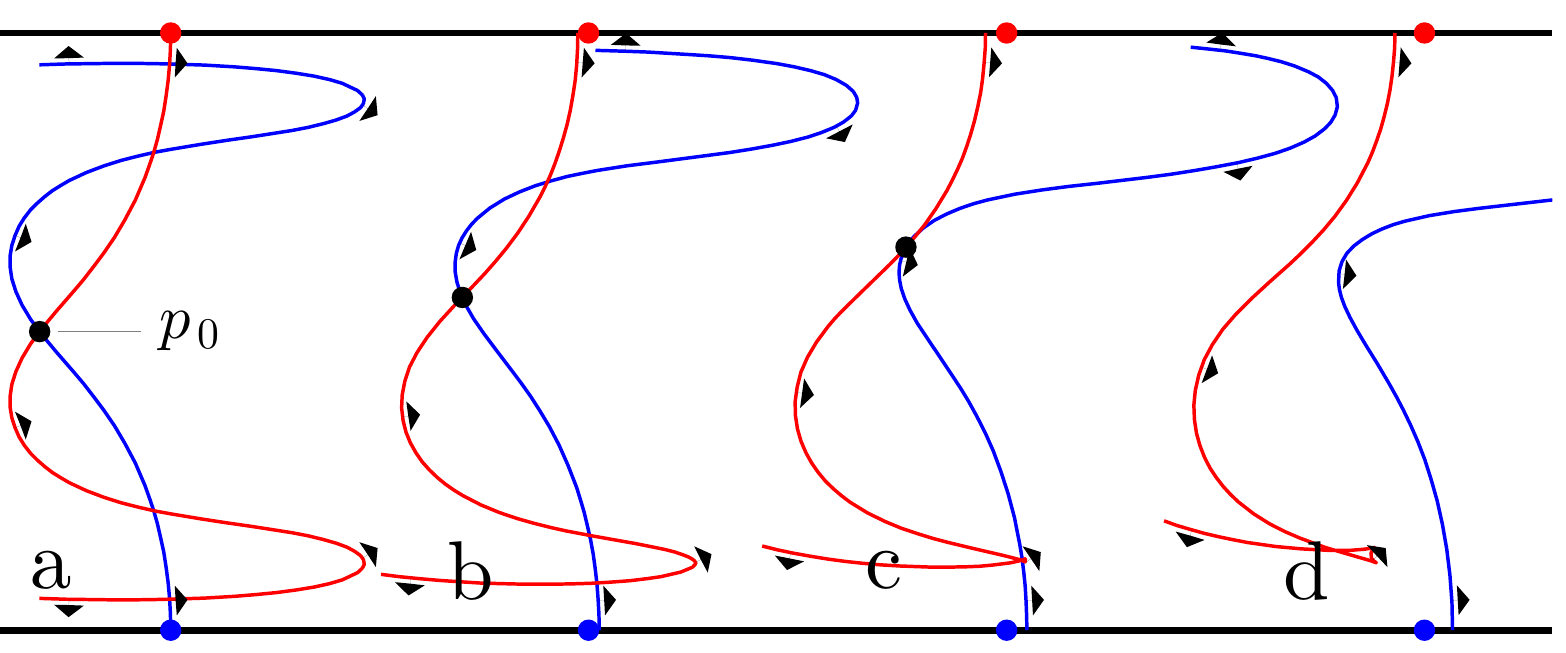}
\caption{Increasing $v_0=$ [0, 0.037, 0.074, 0.1] from left to right. $C_0$ and $C_{-1}$ lobes become empty and burning pips are lost. Dots show advective fixed points.  $(b=0.4, \omega=5, \phi=\pi)$}
\label{fig:v0_wipes_out_lobes}
\end{figure}

Unlike advective pips, which exist for any amount of periodic perturbation, it is possible for burning pips to be pulled apart in certain parameter regimes---for instance, when the front propagation speed is large enough. We can understand the basic intuition geometrically. Figure~\ref{fig:v0_wipes_out_lobes} shows BIMs for the same flow with four increasing values of $v_0$. In the left-most case ($v_0=0$), the BIMs degenerate to the advective invariant manifolds. All four turnstile lobes have the same area. This is because of area-preservation of the map (incompressibility of the fluid) and zero-net-flux across the turnstile. As $v_0$ is increased, the stable and unstable BIMs of interest are to the left and right of their respective advective fixed points. Also, the distance from each fixed point increases, squeezing the capture lobes to zero area, and finally destroying the central pip. Note that this loss of area is a mechanism distinct from the swallowtail discussed above.

It is also interesting to see the heteroclinic connection in Fig.~\ref{fig:v0_wipes_out_lobes}(c). As previously discussed, intersections of fronts in $xy$-space are not intersections in $xy\theta$-space. In this case, since $p_0$ is a tangency between BIM projections, this is a true intersection of BIMs and thus $p_0$ generates a heteroclinic orbit. 

\begin{figure}
\includegraphics[width = \linewidth]{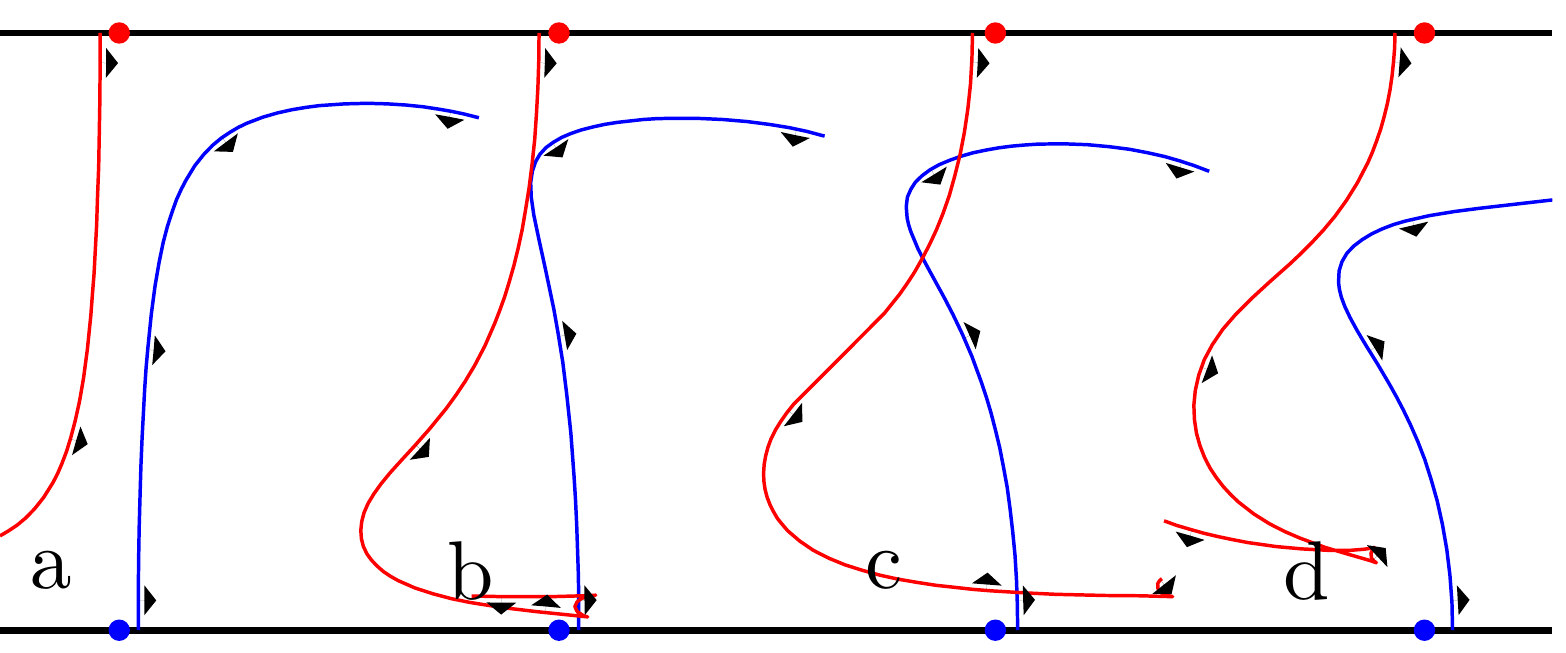}
\caption{Increasing $B=[0.133, 0.267, 0.400, 0.533]$ we see the finite gap caused by $v_0$ overcome by oscillations of the BIMs. Greater $B$ does not necessarily correspond to larger lobes. Dots show advective fixed points. $(v_0=0.1, \omega=5, \phi=\pi)$}
\label{fig:forcing_overcomes_v0_gap}
\end{figure}

We can understand the pip existence issue from another perspective. In the time-independent flow the stable and unstable BIMs are separated by some finite gap (see Fig.~\ref{fig:forcing_overcomes_v0_gap}(a) and Appendix~\ref{appx:no_time_indep_lobes} for details). Sufficient forcing causes oscillations in the BIMs on this same length scale---enough to generate intersections and lobes. The dependence of pip existence and lobe area on forcing strength appears to be more complex than on burning speed. For instance in Fig.~\ref{fig:forcing_overcomes_v0_gap} we see that the pips gained by some forcing in (c) are then lost by greater forcing in (d).

In short, there are several distinct ways in which a system may gain or lose a burning pip. The physical implications of this warrant further investigation.




\subsection{Channel wall}

In this paper, we have considered fluids within a channel geometry. A stimulation made in this channel with no flow will propagate outwards as a circle with radius growing at speed $v_0$ until it runs into the channel wall. A burned region that encounters the channel wall finds nothing beyond to burn, and so its bounding front stops progress in that direction.

While this is straightforward to understand, its implementation is worth noting. Numerically, we supply the channel with a thin boundary layer, extending the domain just beyond the channel wall. The particular flow $\mathbf{u}$ chosen beyond the boundary is not important. 
Upon entering this boundary layer, a front element's velocity perpendicular to the channel is quickly attenuated, reaching zero at the outer limit. In this way, front elements that enter the boundary layer do not leave.

\begin{figure}
\begin{center}
\includegraphics[width=\linewidth]{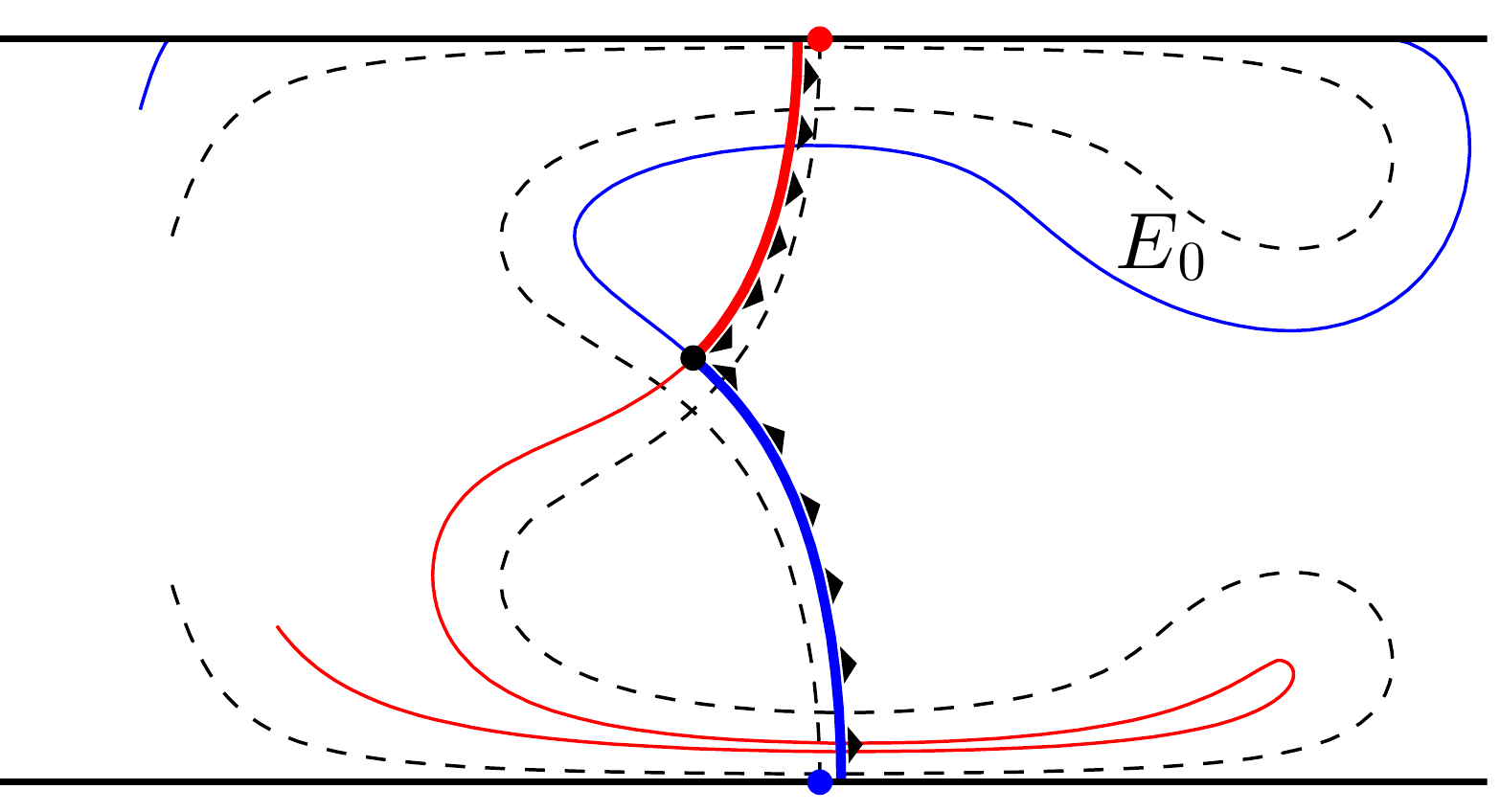}
\caption{The deviation of the BIMs from the advective manifolds (dashed) is sufficient that the unstable BIM runs into the channel wall. The channel wall thus forms part of the $E_0$ lobe boundary. $(v_0=0.075, b=0.3, \omega=4, \phi=\pi)$}
\label{fig:channel_wall_BC_off}
\end{center}
\end{figure}

In the ideal limit the boundary layer is infinitely thin, and front elements that enter this layer can be identified with points on the channel wall.
The result is that when lobes (such as the $E_0$ lobe in Fig.~\ref{fig:channel_wall_BC_off}) collide with the channel wall, the channel wall itself forms a segment of the lobe boundary.

%

\section{Application: reaction front suppression}
\label{sec:reaction_suppression}

There is growing interest in the behavior of biological and chemical systems in fluid flows \cite{Tel05, Neufeld09}. For instance, recent studies have examined the interplay between rapid population growth in plankton and the underlying ocean currents \cite{Sandulescu08}.
Here we describe a scenario in which knowledge of the burning turnstile can be used to more efficiently mitigate the impact of an impinging front. 
Imagine an ocean bay that we wish to protect from an invading algal bloom. Suppose that the surface ocean flow is well-modeled by a 2D periodically perturbed flow \cite{Shadden09}. A separatrix may be generated by an ocean current parallel to the coastline and across the mouth of the bay. Additionally, diurnal tides are a reasonable source of periodic perturbation. This combination produces a turnstile mechanism for ocean water to enter and exit the bay.

As we have discussed in this paper, this turnstile may be modified to account for the addition of propagating fronts. An active constituent in the ocean, such as plankton, would be transported into the bay by the advective turnstile while simultaneously growing outward, pushing past the advective invariant manifolds to the BIMs defining the burning turnstile.

Suppose the goal is to protect the bay by applying an algaecide to the surface of the water in synchrony with the tide. Furthermore, imagine that this substance is 100\% effective, but is active for only a short time, possibly due to photo-degradation. Importantly, treatment  is often costly or toxic to other ocean life, so we wish to apply only as little as possible.
We do not presume the details accurately reflect a real-life scenario.
The intention is to illustrate a plausible motivation for suppressing the invasion of a front.

\subsection{Strategies}

There are several strategies that may be suggested.
First is to apply a strip of algaecide spanning the mouth of the bay in a rectangle. 
While this protocol and simple to implement, it does not take advantage of our understanding of the dynamical structure, and as a result either fails to prevent the algae invasion, or uses an excess of suppressant.

A better strategy attempts to make use of oceanic flow structures. In particular, one could focus the treatment on the advective $E_{-1}$ or $E_0$ lobe. Enlarging this treatment area uniformly around the advective lobe yields an effective and more efficient strategy. As a first approximation, one could enlarge outward a distance $v_0 T$. While this is a reasonable suggestion, it fails to account for the fact that as the front evolves, its continuous propagation beyond the corresponding passive trajectories places the front into regions of flow different from those experienced by the passive trajectories. In other words, the gradient of the flow feeds back into the evolution of the front and in this way leads to lobes that differ from simple uniform enlargements. During the course of one tidal cycle, the algae's growth pushes it just beyond an advective stable manifold. This grants the algae access to a region of fluid that was entirely inaccessible to passive trajectories. This small breach can then proliferate into a significant bloom.

We propose that the best strategy, given the assumptions, is to apply the algaecide to an burning escape lobe $E_{-1}$ or $E_0$. This protocol eliminates exactly that algae which may transgress or has transgressed the transport boundary. Which of the two is the most efficient (smallest area) depends on the details of the flow.

\begin{figure*}
\includegraphics[width=\linewidth]{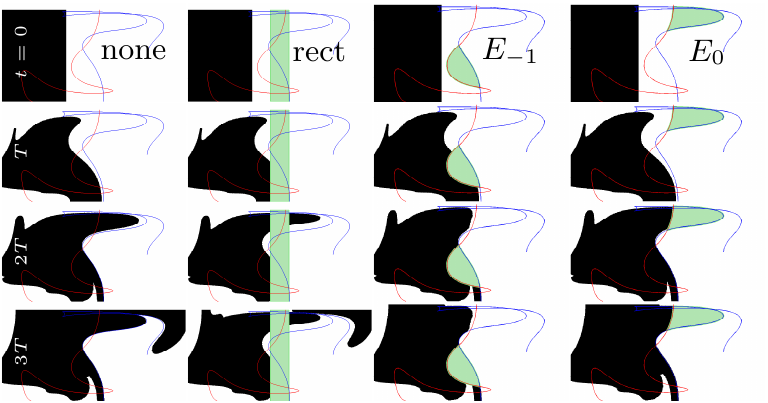}
\caption{Initial algae population (black) impinging from left. Algae evolved using a grid-based algorithm as in Ref.~\cite{Abel01}. Columns show four scenarios. Time increases down. (a) No suppression, algae invades. (b) Rectangular suppression of area $2 |E_0|$ (green) around separatrix. The algae still invades. (c,d) Lobe-based suppression at $E_{-1}, E_0$. Lobe protocols completely prevent invasion.}
\label{fig:reaction_suppression}
\end{figure*}

Figure~\ref{fig:reaction_suppression} illustrates these three protocols (and control with no suppression).
The rectangular protocol clearly fails to prevent the rightward propagation of the front, despite the rectangular area being fully twice that of the larger escape lobe $E_0$.  In contrast, both lobe-based protocols are able to prevent the front propagation, using significantly less suppressant.  (Note that the discrete grid introduces some noise, which we accommodate by increasing the suppression regions by about 2 cell widths.)

For advective transport, one can readily show that the $TB$ defined by a pip connecting stable and unstable manifolds
minimizes the flux across the $TB$.  A small perturbation to such a $TB$ results in an advective flux that is larger than (or equal to) the original flux.  We conjecture that an analogous statement is true concerning the $TB$ for front propagation.  This question, however, is more subtle than in the passive case and will be addressed in a future publication.
%

\section{Summary}
\label{sec:summary}

We have demonstrated how the turnstile, a fundamental construct in the theory of passive fluid transport, is generalized to fronts propagating in a flowing fluid. This burning turnstile is originally defined in terms of the evolution of point stimulations in the active fluid.  The modified turnstile exhibits several features distinct from its passive counterpart. First, the burning turnstile is bounded by BIMs, \emph{one-sided} barriers to front propagation. Second, to form a meaningful transport boundary from finite segments of stable and unstable BIMs, the BIMs relative orientations must be consistent.  Third, the burning pips used to join the stable and unstable BIMS are only intersections of the BIMs when projected to the $xy$ plane.  Fourth, this implies that a segment of the boundary of either $C_{-1}$ or $C_0$ does not consist of BIM segments, depending on whether the pip is concave or convex, respectively.  Finally, the areas of the lobes are not preserved and the relations between $\Masii(E_{-1})$ and $E_0$ and between $\Masii(C_{-1})$ are $C_0$ are more complicated than the advective case.


Increasing the front propagation speed can change the topology of the lobes, most notably by generically introducing swallowtails into the lobe boundaries.  These swallowtails excise a portion of a lobe's BIM boundary.   A large enough propagation speed can even remove a burning pip entirely.  If no such pips remain, the turnstile formulation can no longer be applied.


To illustrate the utility of this theoretical framework, we applied it to a simple model of an ocean bay invaded by algae. This shows that efficient protection of the bay depends on the geometric constructs discussed herein---the burning lobes.

\section{Acknowledgements}

  The authors gratefully acknowledge discussions with Tom Solomon, whose experimental work inspired this paper.  We also benefited from a talk of Solomon's in which he proposed an earlier and alternative formulation of lobes in chaotic advection-reaction-diffusion systems (unpublished).  The present work was supported by the US National Science Foundation under grants PHY-0748828 and CMMI-1201236.


\begin{thebibliography}{10}

\bibitem{Paoletti05}
M.~S. Paoletti and T.~H. Solomon.
\newblock Experimental studies of front propagation and mode-locking in an
  advection-reaction-diffusion system.
\newblock {\em Euro. Phys. Lett.}, 69:819, 2005.

\bibitem{Schwartz08}
M.~E. Schwartz and T.~H. Solomon.
\newblock Chemical reaction fronts in ordered and disordered cellular flows
  with opposing winds.
\newblock {\em Phys. Rev. Lett.}, 100:028302, Jan 2008.

\bibitem{Bargteil12}
Dylan Bargteil and Tom Solomon.
\newblock Barriers to front propagation in ordered and disordered vortex flows.
\newblock {\em Chaos}, 22, 2012.

\bibitem{Pocheau06}
A.~Pocheau and F.~Harambat.
\newblock Effective front propagation in steady cellular flows: A least time
  criterion.
\newblock {\em Phys. Rev. E}, 73:065304, Jun 2006.

\bibitem{Pocheau08}
A.~Pocheau and F.~Harambat.
\newblock Front propagation in a laminar cellular flow: Shapes, velocities, and
  least time criterion.
\newblock {\em Phys. Rev. E}, 77:036304, Mar 2008.

\bibitem{Abel01}
M.~Abel, A.~Celani, D.~Vergni, and A.~Vulpiani.
\newblock Front propagation in laminar flows.
\newblock {\em Phys. Rev. E}, 64:046307, Sep 2001.

\bibitem{Cencini03}
M.~Cencini, A.~Torcini, D.~Vergni, and A.~Vulpiani.
\newblock Thin front propagation in steady and unsteady cellular flows.
\newblock {\em Phys.~Fluids}, 15:679--688, 2003.

\bibitem{Wiggins92}
S.~Wiggins.
\newblock {\em Chaotic Transport in Dynamical Systems}.
\newblock Springer-Verlag, New York, 1992.

\bibitem{Mahoney12}
John Mahoney, Dylan Bargteil, Mark Kingsbury, Kevin Mitchell, and Tom Solomon.
\newblock Invariant barriers to reactive front propagation in fluid flows.
\newblock {\em Euro.~Phys.~Lett.}, 98:44005, 2012.

\bibitem{Mitchell12b}
Kevin~A. Mitchell and John Mahoney.
\newblock Invariant manifolds and the geometry of front propagation in fluid
  flows.
\newblock {\em Chaos}, 22, 2012.

\bibitem{MacKay84}
R.~S. MacKay, J.~D. Meiss, and I.~C. Percival.
\newblock Transport in hamiltonian systems.
\newblock {\em Physica D}, 13:55, 1984.

\bibitem{Solomon88}
T.~H. Solomon and J.~P. Gollub.
\newblock Chaotic particle transport in time-dependent {R}ayleigh-{B}\'enard
  convection.
\newblock {\em Phys. Rev. A}, 38:6280--6286, Dec 1988.

\bibitem{Rom-Kedar90b}
V.~Rom-Kedar and S.~Wiggins.
\newblock Transport in two-dimensional maps.
\newblock {\em Archive for Rational Mechanics and Analysis}, 109:239, 1990.

\bibitem{Note1}
It should be pointed out that the turnstile mechanism is a piece of the larger
  theory of lobe dynamics. The goal of this paper is to firmly establish the
  turnstile mechanism for propagating fronts. A treatment of the
  multi-time-step dynamics is left to a future paper.

\bibitem{Ronney94}
P.~D. Ronney.
\newblock Some open questions in premixed turbulent combustion.
\newblock In J.~Buckmaster, editor, {\em Modeling in Combustion Science}.
  Springer-Verlag, Berlin, 1994.

\bibitem{Fisher37}
R.~A. Fisher.
\newblock The wave of advance of advantageous genes.
\newblock {\em Ann. Eugenics}, 7:355, 1937.

\bibitem{Kolmogorov37}
A.~N. Kolmogorov, I.~G. Petrovskii, and N.~S. Piskunov.
\newblock A study of the diffusion equation with increase in the amount of
  substance, and its application to a biological problem.
\newblock {\em Moscow Univ. Bull. Math.}, 1:1, 1937.

\bibitem{Williams85}
F.~A. Williams.
\newblock {\em Combustion theory : the fundamental theory of chemically
  reacting flow systems}.
\newblock Addison-Wesley, Redwood City, CA, 1985.

\bibitem{Oberlack10}
M.~Oberlack and A.~F. Cheviakov.
\newblock Higher-order symmetries and conservation laws of the g-equation for
  premixed combustion and resulting numerical schemes.
\newblock {\em J.~Eng.~Math.}, 66:121--140, 2010.

\bibitem{Note2}
We say \protect \emph {locally} because unlike the curve in 3D corresponding to
  the perimeter of a compact burned region, a BIM cannot form a topological
  circle.

\bibitem{Rom-Kedar90c}
S.~Wiggins V.~Rom-Kedar, A.~Leonard.
\newblock An analytical study of transport, mixing and chaos in an unsteady
  vortical flow.
\newblock {\em Journal of Fluid Mechanics}, 214:347--394, 1990.

\bibitem{Arnold83}
V.~I. Arnold.
\newblock {\em Catastrophe Theory}.
\newblock Springer, Berlin, 1983.

\bibitem{Note3}
It is important to note that since ray optics is invertible, reversing the
  propagation direction (or time) can cause the swallowtail to unfold. This
  phenomenon is found in the stable BIMs (see Fig.~\ref
  {fig:Cn1swallowtail_w_region}).

\bibitem{Note4}
We are making use of the stability of this perestroika. That is, the structural
  change is actually occurring in time, but according to Ref.~\cite {Arnold83},
  we can see the same morph with a generic parameter sweep (such as $v_0$).

\bibitem{Tel05}
T.~Tel, A.~de~Moura, C.~Grebogi, and G.~Karolyi.
\newblock Chemical and biological activity in open flows: A dynamical system
  approach.
\newblock {\em Phys. Rep.}, 413:91--196, 2005.

\bibitem{Neufeld09}
Z.~Neufeld and E.~Hernandez-Garcia.
\newblock {\em Chemical and Biological Processes in Fluid Flows: A Dynamical
  Systems Approach}.
\newblock Imperial College Press, 2009.

\bibitem{Sandulescu08}
Mathias Sandulescu, Cristobal Lopez, Emilio Hernandez-Garcia, and Ulrike
  Feudel.
\newblock Biological activity in the wake of an island close to a coastal
  upwelling.
\newblock {\em Ecological Complexity}, 5(3):228--237, SEP 2008.

\bibitem{Shadden09}
Shawn~C. Shadden, Francois Lekien, Jeffrey~D. Paduan, Francisco~P. Chavez, and
  Jerrold~E. Marsden.
\newblock The correlation between surface drifters and coherent structures
  based on high-frequency radar data in monterey bay.
\newblock {\em Deep Sea Research Part II: Topical Studies in Oceanography},
  56(3-5):161 -- 172, 2009.

\end{thebibliography}



\appendix

\section{No burning lobes in time-independent fluid flow}
\label{appx:no_time_indep_lobes}
\begin{figure}
\includegraphics[width=\linewidth]{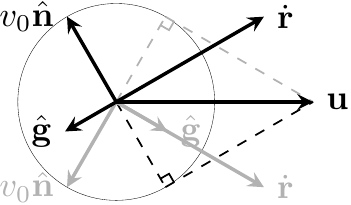}
\caption{For any fluid velocity vector $\mathbf{u}$ where $|\mathbf{u}| > v_0$, there are two solutions (orientations of $\hat{\mathbf{g}}, \hat{\mathbf{n}}$) to the constraint $\mathbf{u} \cdot \hat{\mathbf{n}} + v_0 = 0$. Black solution: $\dot{\mathbf{r}}$, $\hat{\mathbf{g}}$ antiparallel. Gray solution: $\dot{\mathbf{r}}$, $\hat{\mathbf{g}}$ parallel. }
\label{fig:rdot_propto_ghat}
\end{figure}

It is well-known that there are no transverse intersections and hence no advective lobes in time-independent 2D flows.  
These flows also have no burning lobes, however the reasoning behind this is somewhat different.
Assume for the remainder of this section that the fluid flow is time-independent.
As previously shown \cite{Mitchell12b}, BIMs obey the front-compatibility criterion; $\hat{\mathbf{g}} \equiv (\cos \theta, \sin \theta) \propto (dx/d \lambda, dy/d \lambda)$ where $\lambda$ is a parameterization of the BIM by the euclidean distance along the BIM. 
This is a condition on spatial structure.
Additionally, BIMs are, by definition, time-invariant for time-independent flows.
A front element that begins anywhere on the BIM travels in time along the BIM spatially---$(dx/dt, dy/dt) \propto (dx/d\lambda, dy/d\lambda)$.
These two proportionalities combine to yield $\hat{\mathbf{g}} \equiv (\cos \theta, \sin \theta) \propto (dx/dt, dy/dt) = \mathbf{u} + v_0 \hat{\mathbf{n}}$. This implies $\mathbf{u} \cdot \hat{\mathbf{n}} + v_0 = 0$.
This is our main constraint on BIMs in time-independent flows.
If $|\mathbf{u}| < v_0$, there are no solutions.
If $|\mathbf{u}| = v_0$, there is one solution $\hat{\mathbf{n}} =  -\mathbf{u} / v_0$.
If $|\mathbf{u}| > v_0$, there are two solutions which we illustrate in Fig.~\ref{fig:rdot_propto_ghat}.
Consider the (hypothetical) intersection of two BIM segments at a burning pip.
If at this intersection point there is one constraint solution, both BIM segments pass through the same $xy\theta$ point. 
By uniqueness of solutions to ODEs, these trajectories are the same and so there are no burning lobes.
If there are two constraint solutions, the two BIM segments pass through the two different points $(x_0, y_0, \theta_0)$ and $(x_0, y_0, \theta_0')$. If they pass through the same point, the previous argument applies again.
Therefore the intersection must be transverse and the two front elements involved must bear the relation in Fig.~\ref{fig:rdot_propto_ghat}.
Finally notice that a burning pip, such as $p_0$ or $q_0$ in Fig.~\ref{fig:ps_and_qs}), does not have this configuration.
Hence we reach a contradiction.
This demonstrates that time-independent flows do not allow transverse pips or burning lobes.

\end{document}